\documentclass[twocolumn]{aastex62}

\submitjournal{ApJ}

\graphicspath{{images/}}
\usepackage{graphicx}
\usepackage{enumitem}
\usepackage{amsmath}
\usepackage{bm}

\shorttitle{Pair production in relativistic reconnection}
\shortauthors{Hakobyan et al.}

\definecolor{darkgreen}{HTML}{04af70}
\definecolor{darkred}{HTML}{bf0000}

\newcommand{\linkVidOne}{https://youtu.be/VhwXzci_bY8} 
\newcommand{\linkVidTwo}{https://youtu.be/Mu_VYtADbnc} 
\newcommand{\linkVidThree}{https://youtu.be/QY_Bpjr1p2Q} 

\begin{document}

\title{Effects of synchrotron cooling and pair production on collisionless relativistic reconnection}

\correspondingauthor{Hayk Hakobyan}
\email{hakobyan@astro.princeton.edu}

\author{Hayk Hakobyan}
\affil{Department of Astrophysical Sciences, Peyton Hall, Princeton University, Princeton, NJ 08544, USA}

\author{Alexander Philippov}
\altaffiliation{Einstein Fellow}
\affiliation{Department of Astronomy, University of California, Berkeley, CA 94720-3411, USA}
\affiliation{Center for Computational Astrophysics,
Flatiron Institute, New York, NY 10010, USA}

\author{Anatoly Spitkovsky}
\affil{Department of Astrophysical Sciences, Peyton Hall, Princeton University, Princeton, NJ 08544, USA}

\begin{abstract}
High energy radiation from nonthermal particles accelerated in relativistic magnetic reconnection is thought to be important in many astrophysical systems, ranging from blazar jets and black hole accretion disk coronae to pulsars and magnetar flares. The presence of a substantial density of high energy photons ($>$MeV) in these systems can make two-photon pair production ($\gamma\gamma\to e^-e^+$) an additional source of plasma particles and can affect the radiative properties of these objects. We present the results of novel particle-in-cell simulations that track both the radiated synchrotron photons and the created pairs, with which we study the evolution of a two-dimensional reconnecting current sheet in pair plasma. Synchrotron radiation from accelerated particles in the current sheet produces hot secondary pairs in the upstream which are later advected into the current sheet where they are reaccelerated and produce more photons. In the optically thin regime, when most of the radiation is leaving the upstream unaffected, this process is self-regulating and depends only on the background magnetic field and the optical depth of photons to pair production. The extra plasma loading also affects the properties of reconnection. We study how the inflow of the secondary plasma, with multiplicities up to several hundred, reduces the effective magnetization of the plasma, suppressing the acceleration and thus decreasing the high energy photon spectrum cutoff. This offers an explanation for the weak dependence of the observed gamma-ray cutoff in pulsars on the magnetic field at the light cylinder. 

\end{abstract}

\section{Introduction} \label{sec:intro}

One of the key observational characteristics of $\gamma$-ray pulsars is a hard power law spectrum (with a typical photon index $\Gamma\sim 1\text{-}2$) extending to cutoff energies of a few GeV \citep{2013ApJS..208...17A}. While pulsed radio emission is generated in the inner magnetosphere close to the neutron star surface, gamma radiation usually arrives in different rotational phases and based on its light curve properties is thought to be produced mainly in the outer magnetosphere close to the Y-point (see, e.g., \citealt{2010ApJ...715.1282B, 2010MNRAS.404..767C, 2015ApJ...811...63H}). Reconnection in the current layer beyond the Y-point just outside the light cylinder is a plausible mechanism of non-thermal particle acceleration that can later form the observed spectrum via synchrotron radiation \citep{1990ApJ...349..538C, 1996A&A...311..172L, 2012MNRAS.424.2023P, 2016MNRAS.457.2401C, PSAS18}. \par

\begin{figure}[htb]
    \centering
    \includegraphics[width=\columnwidth]{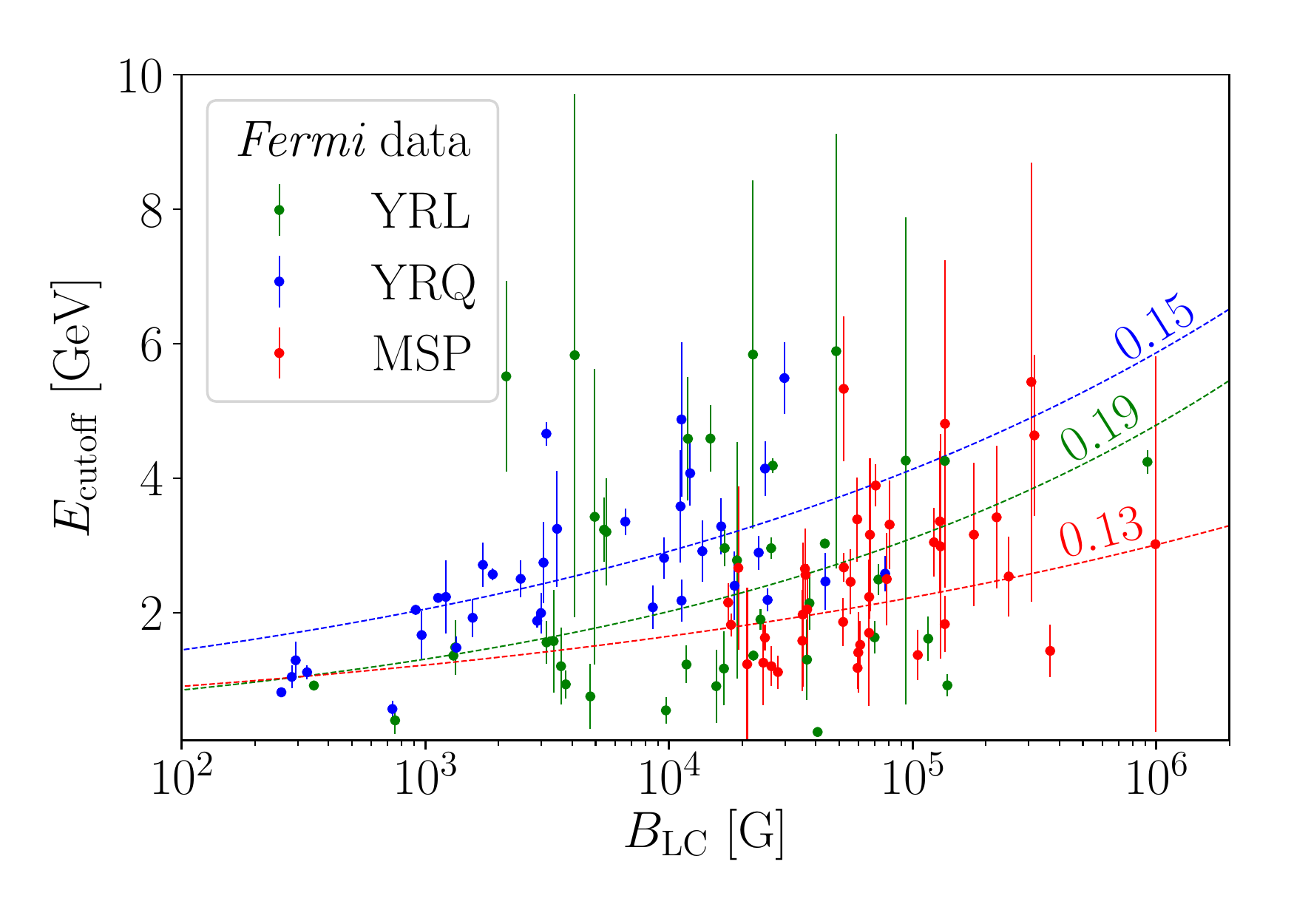}
    \caption{Data from {\it Fermi Observatory} observations of the cutoff energy of photon spectra plotted against the estimated light cylinder magnetic fields for young radio-loud, young radio-quiet, and millisecond pulsars. The dashed lines are the best fit power laws, and the corresponding numbers are the power-law photon indices. Data taken from  \href{https://fermi.gsfc.nasa.gov/ssc/data/access/lat/2nd_PSR_catalog/}{LAT Second Catalog of Gamma-ray Pulsars}.}
    \label{fig:fermi_data}
\end{figure}

However, while the light curves and spectral shapes of pulsar $\gamma$-ray emission can be naturally explained with the current sheet emission, there is a disagreement between the observations and the predictions of spectral cutoff of this model. The high energy spectral cutoff in pulsars, as measured by the {\it Fermi Observatory}, depends weakly on the magnetic field at the light cylinder, $B_{\rm LC}$. As shown in Figure~\ref{fig:fermi_data}, for a wide range of $B_{\rm LC}=10^3\text{-}10^6$ G the cutoff energy varies from $1$ to around $6$ GeV \citep{2013ApJS..208...17A}. On the other hand, the model that incorporates reconnection predicts a stronger dependence. In relativistic reconnection the magnetic energy that can be deposited into particle kinetic energy is controlled by the magnetization parameter at the light cylinder, $\sigma_{\rm LC}$, defined as the ratio of the magnetic energy to plasma enthalpy. For the cold plasma this parameter is equal to: 
\begin{equation}\label{eq:sigma_def}
    \sigma_{\rm LC}\approx \frac{B_{\rm LC}^2/4\pi}{n_{\rm LC}m_e c^2},
\end{equation}
where $B_{\rm LC}$ and $n_{\rm LC}$ are the magnetic field and the plasma density near the light cylinder. In particular, in a strongly magnetized plasma, $\sigma_{\rm LC}\gg 1$, the particle spectrum cutoff does not typically exceed few times $\sigma_{\rm LC}$ \citep{2014ApJ...783L..21S, 2016ApJ...816L...8W}. Note here, that there are mechanisms that can increase this cutoff to much higher energies in the uncooled reconnection \citep{2018arXiv180800966P}. We will discuss later why this is not possible when strong synchrotron cooling is present. 

The plasma loading along the separatrix to the light cylinder and outer magnetosphere is set by the multiplicity of the primary cascade near the polar cap, $\kappa\sim 10^4$, \citep{1982ApJ...252..337D,2013MNRAS.429...20T,2018arXiv180308924T} and the local Goldreich-Julian density, $n_{\rm GJ}\approx\Omega B_{\rm LC}/2\pi c e$, i.e., $n_{\rm LC}\approx \kappa n_{\rm GJ}$. The multiplicity of primary cascade, $\kappa$, is roughly insensitive to the magnetic field strength. This means that the plasma density scales linearly with the magnetic field at the light cylinder, and thus from formula~(\ref{eq:sigma_def}) the maximum particle energy, $\gamma_{\rm max}\sim \sigma_{\rm LC}$, also scales linearly with $B_{\rm LC}$. If these particles radiate synchrotron photons which form the observed $\gamma$-ray emission, the cutoff energy of photon spectrum, $E_{\rm cutoff}$, will correspond to the maximum energy of particle spectrum set by $\sigma_{\rm LC}$. This will lead to a strong dependency of the cutoff energy on the background magnetic field: $E_{\rm cutoff}\propto \gamma_{\rm max}^2 B_{\rm LC}\propto B_{\rm LC}^3$. This discrepancy with the observed weak dependency suggests that there must be a self-regulating source of additional mass loading of the current layer that effectively decreases the magnetization and suppresses the particle acceleration for higher magnetic fields.

In the region close to the current layer the number density of high energy synchrotron photons is sufficiently high for two-photon pair production to be efficient. This process has previously been studied in the context of outer gap acceleration models \citep{1996A&AS..120C..49A, 2010ApJ...715.1318T}. As was shown by \cite{1996A&A...311..172L} for the reconnection-powered acceleration occurring in the equatorial current sheet, this pair creation process can significantly increase the plasma population with multiplicities of secondary pairs up to several thousand ($\sim 3000$ for Crab). On the other hand, thermal keV radiation from the neutron star surface is insufficiently luminous to interact with high energy GeV photons and significantly contribute to pair production. Thus, the main effect is due to high energy (keV to GeV) synchrotron photons emitted locally in the sheet interacting with each other in the current layer near the Y-point. 

Since plasma loading of the reconnection layer affects the rate of reconnection and its acceleration properties, it is reasonable to expect that the system will reach a self-consistent steady state in both particle and radiation spectra. Such steady states driven by secondary pair plasma loading can also have observational manifestations in other environments where extreme reconnection with pair production is thought to occur, such as in blazar jets~\citep{2009MNRAS.395L..29G}, magnetar flares~\citep{1994MNRAS.270..480T}, or black hole disk coronae~\citep{2017ApJ...850..141B}.

In this paper we study the previously unexplored regime of reconnection with self-consistent pair production that achieves high multiplicities of secondary plasma. We present the results of relativistic particle-in-cell simulations of a two-dimensional reconnecting current sheet with photon radiation and pair production, where we create and track photons as separate particles and self-consistently incorporate two-photon pair production events. In \S\ref{sec:predict} we revisit the general picture of relativistic magnetic reconnection and introduce the main parameters of the problem. We then discuss an analytical model of how pair production and secondary plasma loading affects the steady state of reconnection and, as a result, the emerging particle and photon spectra. In \S\ref{sec:setup} we introduce the numerical setup that we used and describe our algorithm. In \S\ref{sec:results} we summarize the main results and discuss how they can be applied to understand the high energy radiation from pulsars.

\section{Reconnection with pair production: theory}
\label{sec:predict}

\begin{figure*}[ht]
    \centering
    \centerline{\includegraphics[width=2.2\columnwidth]{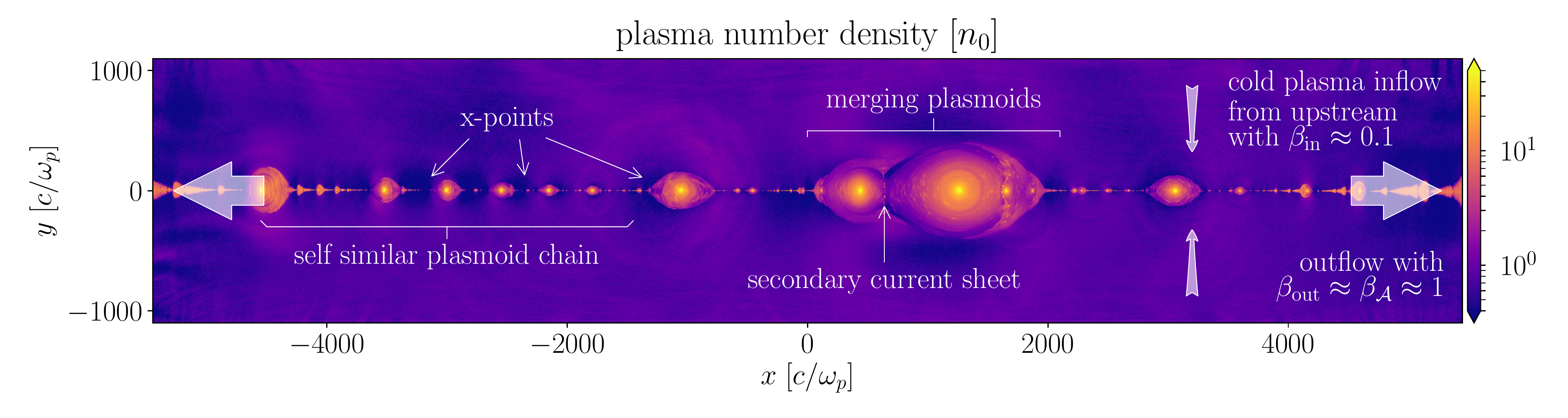}}
    \caption{Reconnection in two-dimensional box with injection from top and bottom and outflowing boundaries on the left and right. The upstream magnetic field is in the plane of the picture. The color represents the plasma density normalized to the upstream value, $n_0$. $\beta=v/c$ is the typical inflow or outflow velocity. The image is taken from an actual simulation.}
    \label{fig:rec_scheme}
\end{figure*}

In this section we give a theoretical outline of our paper as well as introduce the main terminology and current understanding of how relativistic reconnection works.

\subsection{Relativistic reconnection in $e^\pm$ plasma}

It is well known that a current layer separating uniform oppositely directed magnetic field regions is unstable. Plasmoid instability, developing quickly on Alfv\'en crossing timescales, initiates  magnetic reconnection, breaking the current layer into a chain of outflowing plasmoids (see, e.g., \citealt{1997plasconf, 2007PhPl...14j0703L}). Energy of the magnetic field is liberated in this process, which in the relativistic case powers non-thermal particle acceleration, forming a power law spectrum of particles with a cutoff set by the magnetization.\par

In Figure~\ref{fig:rec_scheme} we show a typical simulation snapshot of a two dimensional current sheet undergoing the plasmoid instability and reconnection. All the important features, such as the plasmoids, plasmoid mergers, and secondary current sheets are shown in the figure for guidance. The pattern of plasmoids is self-similar on a wide range of scales, and their evolution is stochastic: smallest plasmoids move along the layer with the Alfv\'en velocity (which in relativistic case is close to $c$), but they also merge to become larger and slower, with bulk motions that are just marginally relativistic (for more details see~\citealt{2016MNRAS.462...48S}). 

Between the plasmoids, the annihilated magnetic field in the x-points allows the reconnection electric field to accelerate particles. Along with the primary current layer, there are also secondary layers which occur during the plasmoid collisions, serving as additional locations of particle acceleration \citep{2015ApJ...806..167G}. Plasmoids advect hot plasma containing the most energetic particles along the sheet with the Alfv\'en velocity, while the cold upstream plasma, having typically small Lorentz-factors, $\langle\gamma\rangle\approx 1$, inflows with a characteristic velocity $\beta_{\rm in}\equiv v_{\rm in}/c\sim0.1\text{-}0.2$, which corresponds to the reconnection rate. \par

The main dimensionless parameter controlling the dynamics of collisionless relativistic reconnection is the cold magnetization parameter, $\sigma_{\rm c}\equiv (B^2/4\pi) / n m_e c^2$, that sets the available magnetic energy per particle. Here, $n$ is the pair plasma density far from the reconnection layer, where the unreconnected field has the value $B$. When $\sigma_{\rm c}\gg 1$, the majority of particles accelerated in the current layer is typically in the ultra-relativistic regime. 
Magnetization controls the maximum energy to which a particle can be accelerated in reconnection. In particular, as was shown by~\cite{2014ApJ...783L..21S} and \cite{2016ApJ...816L...8W}, in the x-point particles can be accelerated up to a few times $\sigma_{\rm c}$. The plasmoids can further accelerate particles, extending the power law to much higher energies \citep{2001ApJ...562L..63Z, 2016MNRAS.462...48S, 2016ApJ...818L...9G, 2018arXiv180800966P}. 

In pulsars the value of magnetization, which 
depends both on the magnetic field strength and the plasma density, can be estimated from the spin down power of the pulsar and the properties of the initial pair production cascade near the polar caps~\citep{1983AIPC..101..163A, 1995phpu.book.....L, 2001ApJ...547..437L, 2010MNRAS.406.1379M, 2010MNRAS.408.2092T}. Typically, this parameter lies between $10^3\text{-}10^5$; for the Crab it can also be directly constrained from PWN observations to be close to $10^4$. This value sets the cutoff energies to which particles can be accelerated. 



\subsection{Particle cooling and photons}
\label{subsec:pcoolandphot}

Both electrons and positrons are subject to radiation cooling at relativistic energies. In this paper we only consider synchrotron cooling, while  inverse-Compton (IC) cooling may also be important in some contexts~\citep{2018arXiv180501910W}. If the synchrotron cooling time is shorter than the acceleration time (we call this a strong cooling regime), accelerated particles will quickly lose their gained energy in the large background magnetic field without having a chance to reaccelerate again. Unlike the case without cooling, the highest energy particles will no longer be located within the plasmoids, since the plasmoids have typically higher magnetic fields, and the particles within them are efficiently cooled. Instead, most of the high energy particles will be piled up in the vicinity of plasmoids and along the primary and secondary current layers where the magnetic field is close to zero; the maximum energy will thus no longer be set by the plasmoid sizes. Even when synchrotron cooling is strong, the particle energy cutoff will be close to a few times $\sigma_{\rm c}m_e c^2$, as these particles would still accelerate in the primary and secondary current sheets where the local cooling is inefficent.


 
The cooling regime is parametrized by the value of the Lorentz-factor of particles, $\gamma_{\rm rad}$, for which the radiation drag force is comparable to the accelerating force (for the accelerating electric field we assume $E \sim \beta_{\rm rec} B_0$; hereafter, we use subscript ``0" for upstream values):
\begin{equation}
    \label{eq:def_gammarad}
    2\sigma_{\rm T}\frac{B_0^2}{8\pi}\gamma_{\rm rad}^2 = e\beta_{\rm rec}B_0,
\end{equation}
where $\beta_{\rm rec}\approx 0.1$ is the steady-state reconnection rate, and $\sigma_{\rm T}$ is the Thomson cross section. 


The radiation from a single plasma particle is described by the synchrotron spectrum, peaking at frequency $\omega_{\rm syn}\approx eB_0\gamma^2 / m_e c$. An important benchmark energy for pair production is the electron (positron) rest-mass energy, $m_e c^2$, which determines the minimum center-of-momentum energies for two photons to pair produce. We are, thus, interested to know which plasma particles radiate photons with characteristic energies close to $m_e c^2$. This sets another dimensionless parameter -- the Lorentz-factor of these particles, $\gamma_c$, determined by
\begin{equation}
    \label{eq:def_gammac}
    \hbar \frac{e B_0 \gamma_c^2}{m_e c}= m_e c^2.
\end{equation}

Combined together, the cold magnetization parameter of the upstream, $\sigma_{\rm c}$, radiation-reaction limit, $\gamma_{\rm rad}$, and the pair threshold parameter, $\gamma_c$, give the full description of the synchrotron-cooled reconnection problem. We can rewrite the definitions as
\begin{equation}
\label{eq:def_gammas}
    \gamma_{\rm rad}^2 \equiv \frac{3\beta_{\rm rec}}{2}\frac{B_{\rm cl}}{B_0},~~~
    \gamma_c^2 \equiv \frac{\alpha B_{\rm cl}}{B_0} = \frac{B_{\rm S}}{B_0},
\end{equation}
where $\alpha$ is the fine-structure constant, $1/137$, $B_{\rm cl}=m^2 c^4/e^3$ is the classical magnetic field, and $B_{\rm S}=m_e^2 c^3/e\hbar$ is the Schwinger field.

For a typical pulsar with the magnetic field at the light cylinder $B_0 = B_{\rm LC} \sim 10^{5}$ G, we find
\begin{equation}\label{eq:rad_c_values}
    \gamma_{\rm rad} \approx 10^5 \left(\frac{B_0}{10^{5}~\textrm{G}}\right)^{-1/2},~~~
    \gamma_c \approx 2\cdot 10^4 \left(\frac{B_0}{10^{5}~\textrm{G}}\right)^{-1/2}.
\end{equation}
For the Crab, with $B_{\rm LC}\sim 4\times 10^6$ G, these values are $\gamma_{\rm rad}\approx 10^4$ and $\gamma_c\approx 3\times 10^3$~\citep{2014ApJ...780....3U}, and the typical magnetization near the light cylinder is $10^4\text{-}10^5$. We, thus, have a hierarchy of energy scales with $\gamma_c \ll \gamma_{\rm rad} \lesssim \sigma_{\rm c}$, which we will use in our simulations.

\subsection{Two-photon pair production}
\label{subsec:twophotpp}

\begin{figure}[tb]
    \centering
    \includegraphics[width=\columnwidth]{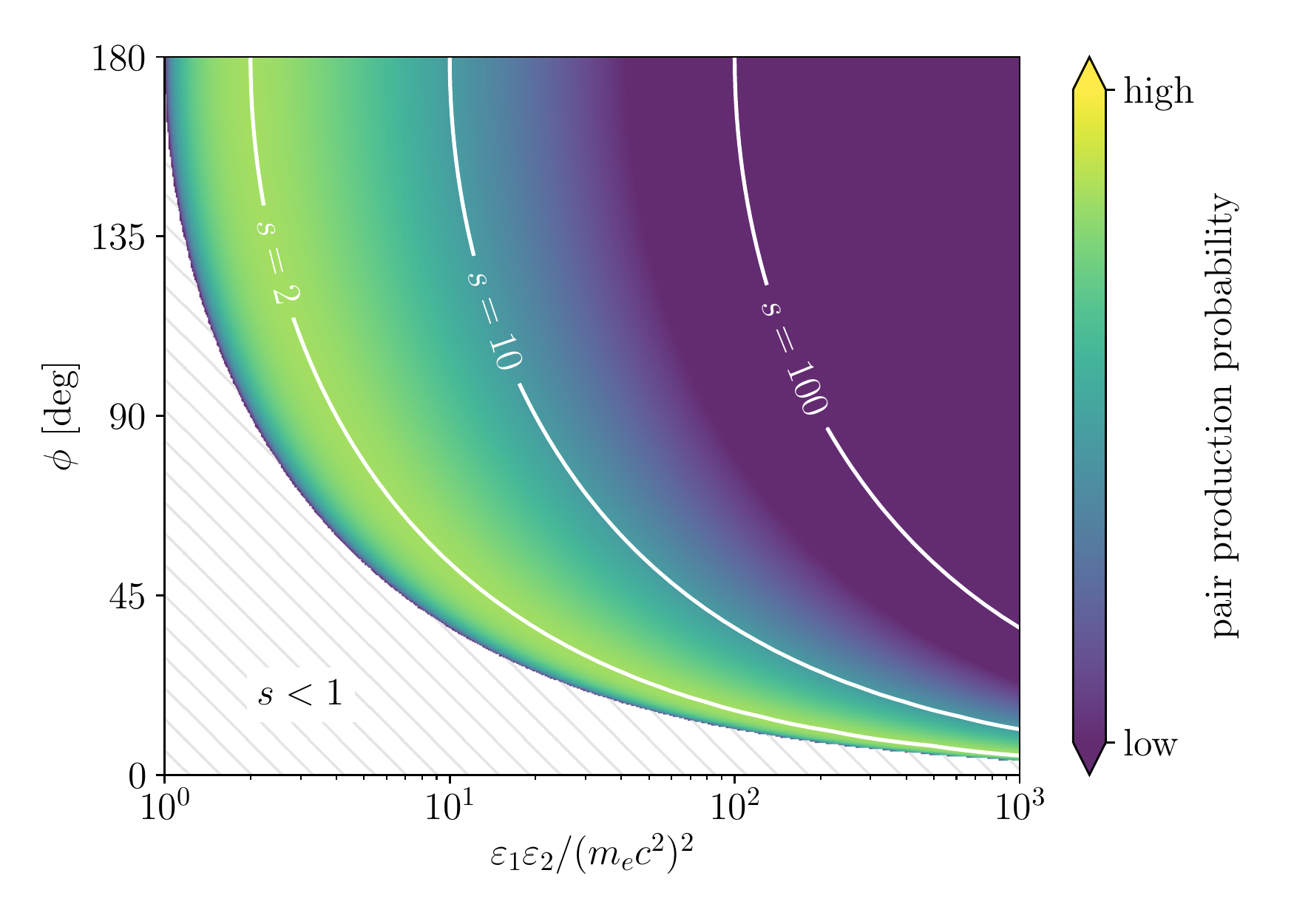}
    \caption{Two-photon pair production cross section as a function of the angle between their momenta, $\phi$, and the product of photon energies in units of $m_e c^2$. The pair production probability is highest when the center-of-momentum energy is $\approx \sqrt{2}m_e c^2$. For small angles only the most energetic photons can pair produce with each other, while head-on interactions allow a wider range of energies.}
    \label{fig:sigma_pp}
\end{figure}

Two photons can interact through the Breit-Wheeler process to form an electron-positron pair, $\gamma\gamma\to e^-e^+$ \citep{PhysRev.46.1087}. This can happen if the center-of-momentum energy of photons is greater than the rest-mass energy of the electron-positron pair
\begin{equation}
    \label{eq:def_s}
    s \equiv \frac{1}{2}\frac{\varepsilon_1\varepsilon_2}{(m_e c^2)^2} (1-\cos{\phi}) > 1,
\end{equation}
where $\varepsilon_1$ and $\varepsilon_2$ are the lab frame photon energies, and $\phi$ is the angle between their momenta. The cross section for this interaction behaves as $\sqrt{s-1}$ near $s\gtrsim 1$, peaks around $s\approx 2$ and drops down as $1/s$ for $s\gg 1$. In Figure~\ref{fig:sigma_pp} we show the magnitude of this cross section plotted vs the relative angle of two interacting photons, $\phi$, and the product of their energies measured in $m_ec^2$. White shaded region corresponds to values of $s$ where pair production is not possible.

Figure \ref{fig:sigma_pp} demonstrates two important facts: the high energy photons ($\varepsilon \gg m_e c^2$) pair produce preferentially with the lower energy ones ($\varepsilon \ll m_e c^2$), while the ones with intermediate energies ($\varepsilon \approx m_e c^2$) pair produce with each other. Also, the small angle interactions with $\phi\approx 0$ are suppressed, while the head-on collisions, $\phi\approx \pi$, are more preferred.

We will consider a system to be optically thin to two-photon pair production, $\tau_{\gamma\gamma}\ll 1$, if at all energies only a small fraction of photons is converted to pairs. This means that most of the photons stream freely out of the system without any interactions. Note also, that this condition is hardest to satisfy for the highest energy photons, since they typically have a high pair production probability while streaming through a dense background of low energy radiation.

In an optically thin regime for a power-law energy distribution of photons one can show that photons in a wide range of energies contribute roughly equally to pair production. If the photon energy cutoff is at $\varepsilon_{\rm max}$, all the photons from $\varepsilon_{\rm min} \approx (m_e c^2)^2/\varepsilon_{\rm max}$ to $\varepsilon_{\rm max}$ are equally important to consider. This makes the problem of photon tracking numerically challenging. On the other hand, very low energy photons, $\varepsilon \ll \varepsilon_{\rm min}$, have no high energy partner to interact with, and can thus be thrown out of consideration in the context of pair production.

\subsection{Steady state and the effective secondary plasma density}
\label{sec:theory}

\begin{figure}[tb]
    \centering
    \includegraphics[width=\columnwidth]{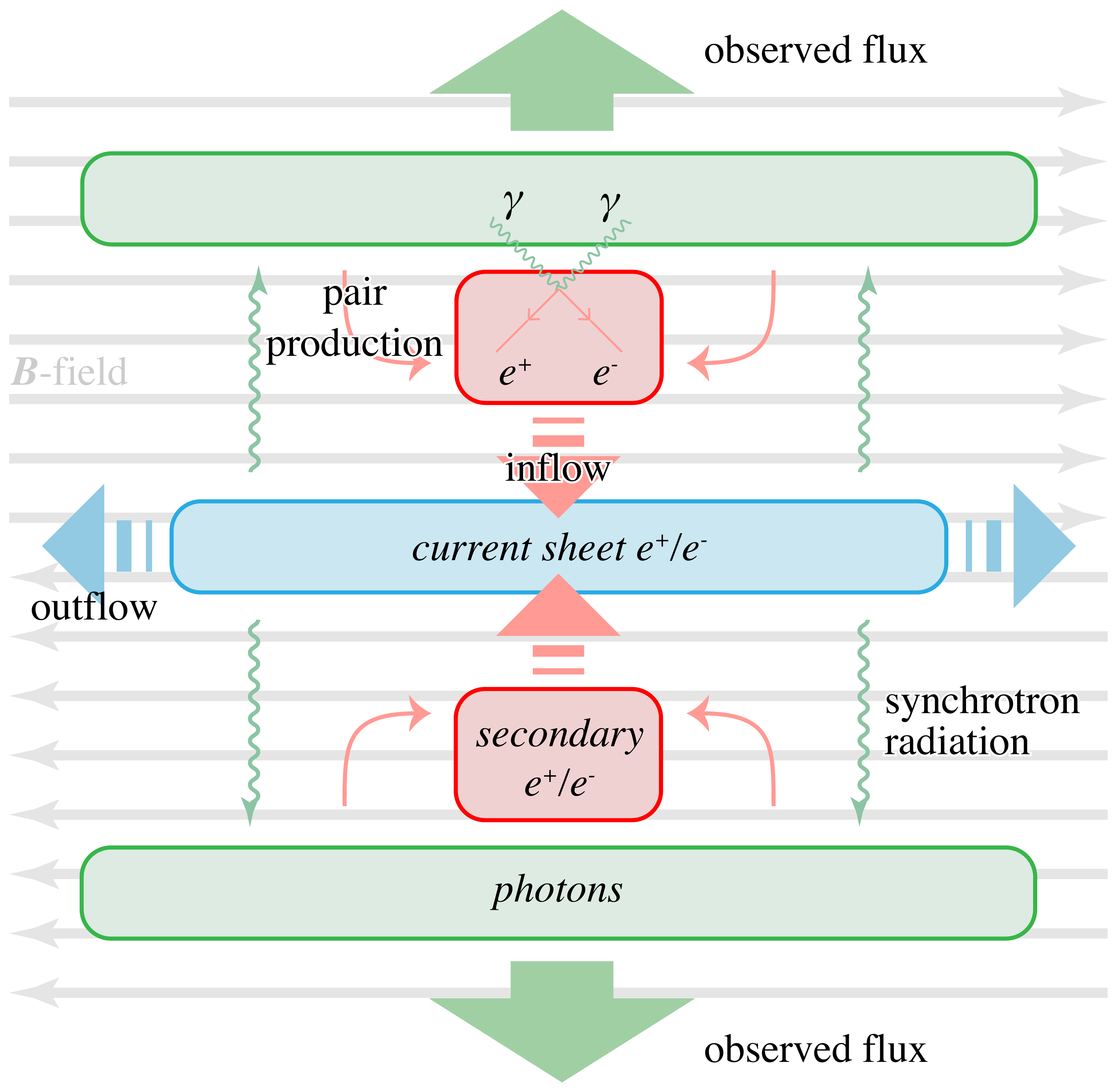}
    \caption{Schematic illustration of the reconnecting current sheet in its steady state. Plasma in the current sheet cools down via synchrotron radiation. These photons later pair produce in the upstream, and secondary pairs are advected into the current sheet.}
    \label{fig:steady_state}
\end{figure}

We now describe a simple model of reconnection with synchrotron cooling and pair production to better illustrate the pair loading feedback mechanism and the steady state. We also predict the steady state multiplicity of secondary plasma in our simulations and in $\gamma$-ray pulsars, and it's dependence on the background magnetic field.

The general picture is schematically shown in Figure~\ref{fig:steady_state}, where we show how different ingredients of reconnection work together to build up the feedback and ultimately drive the system to a steady state. Most of the plasma resides in the current sheet (blue region in Fig.~\ref{fig:steady_state}), where it gets accelerated and radiates a power law distribution of photons $\mathrm{d}n_\gamma/\mathrm{d}\varepsilon_\gamma\propto\varepsilon_\gamma^{-\Gamma}$. Each particle experiences a radiative drag force and loses its energy to radiation at the rate
\begin{equation}
    \label{eq:sync_rate}
    \dot{\epsilon} = 2\sigma_{\rm T} c \frac{B_0^2}{8\pi}\gamma^2 = e c \beta_{\rm rec} B_0 \left(\frac{\gamma}{\gamma_{\rm rad}}\right)^2,
\end{equation}
where $\gamma_{\rm rad}$ is defined in Eq.~(\ref{eq:def_gammarad}).

We will assume that each particle with energy $\gamma m_e c^2$ radiates photons with synchrotron peak frequency in the fixed upstream field $B_0$:
\begin{equation}
    \label{eq:sync_peak}
    \varepsilon_\gamma = m_e c^2\left(\frac{\gamma}{\gamma_c}\right)^2,
\end{equation}
where $\gamma_c$ is defined in Eq.~(\ref{eq:def_gammac}). The radiated photons pair produce in the upstream, creating secondary pairs that feed the current sheet further. The total photon production rate can be found from  equations (\ref{eq:sync_rate}) and (\ref{eq:sync_peak}):
\begin{equation}\label{eq:photon_rate}
    \dot{n}_\gamma = \frac{\dot{\epsilon} (n_{\rm sec} + n_0)}{\varepsilon_\gamma} = \frac{e c \beta_{\rm rec} B_0 n_0 (1+\eta)}{m_e c^2}\left(\frac{\gamma_c}{\gamma_{\rm rad}}\right)^2,
\end{equation}
where we defined the multiplicity, $\eta$, as the ratio of the secondary plasma density to that of the primary, $n_{\rm sec} / n_0$.

Given the spectrum of photons we can find how the number of pair producing photons scales with the spectral cutoff
\begin{equation}\label{eq:photon_pp_scale}
    \tilde{n}_\gamma \propto \int\limits_{\varepsilon_{\rm min}}^{\varepsilon_{\rm max}}\varepsilon_\gamma^{-\Gamma}\mathrm{d}\varepsilon_\gamma \propto \varepsilon_{\rm min}^{1-\Gamma}\propto \varepsilon_{\rm max}^{\Gamma - 1},
\end{equation}
where we used the fact, that $\varepsilon_{\rm min}\varepsilon_{\rm max}\approx (m_e c^2)^2$.


Photons of the highest energies are radiated by the particles with the highest Lorentz-factors. Reconnection accelerates particles up to a few times $\sigma$, which is lower than the initial value $\sigma_0=B_0^2/4\pi n_0 m_e c^2$ by a factor of $(1+\eta)$. Thus, when the multiplicity grows, the effective magnetization and the maximum photon energy decrease, shrinking the energy band of pair producing photons (decreasing $\tilde{n}_{\gamma})$. This is the pair production feedback that drives the system to a self-regulated steady state.

To estimate the steady state multiplicity we can write the pair production rate as follows:
\begin{equation}\label{eq:pair_rate}
    \dot{n}_{\rm sec} \approx \sigma_{\rm T} c \tilde{n}_\gamma^2 f_0,
\end{equation}
where the dimensionless parameter $f_0$ sets the interaction cross section, $\sigma_{\gamma\gamma} = f_0 \sigma_{\rm T}$, and depends on the energies and momenta of interacting photons. The peak value of $f_0$ is close to $0.3$; we will take an empirical value of $0.1$. The final result depends weakly on this parameter.

In the steady state both the secondary plasma and the photons are advected out of the reconnection region. We introduce two length-scales, $s_1$ and $s_2$, which define the size of the region where most of the photon radiation and pair production take place. We can then transform from the production rates to steady state densities: 
\begin{equation}\label{eq:def_scales}
    n_\gamma = \dot{n}_{\gamma} s_1 / c,~~~n_{\rm sec} = \dot{n}_{\rm sec} s_2 / c.
\end{equation}
Combining Eq.~(\ref{eq:photon_rate}-\ref{eq:def_scales}) we find
\begin{equation}
\begin{split}
    n_{\rm sec} = n_0 \eta & = 
                    \beta_{\rm rec}^2 f_0 \frac{\sigma_{\rm T} e^2 B_0^2 n_0^2}{(m_e c^2)^2} s_1^2 s_2 \times \\
                    & \times (1+\eta)^2\left(\frac{\varepsilon_{\rm max}}{m_e c^2}\right)^{2\Gamma-2}\left(\frac{\gamma_c}{\gamma_{\rm rad}}\right)^4,
\end{split}
\end{equation}
where we set $\tilde{n}_\gamma = n_{\gamma}(\varepsilon_{\rm max}/m_e c^2)^{\Gamma-1}$. We can then take $\varepsilon_{\rm max}/m_e c^2 = (\sigma/\gamma_c)^2$, with $\sigma = \sigma_0 / (1 + \eta)$, and substitute $n_0 = B_0^2 / 4\pi \sigma_0 m_e c^2$. After some simplification we get
\begin{equation}
    \frac{\eta}{(1+\eta)^{6-4\Gamma}} = 
            \frac{\beta_{\rm rec}^2 f_0}{4\pi}
            \frac{\sigma_{\rm T} e^2 B_0^4}{(m_e c^2)^3}s_1^2 s_2
            \frac{\gamma_c^{8-4\Gamma}}{\gamma_{\rm rad}^4\sigma_0^{5-4\Gamma}}.
\end{equation}

To express this relation through dimensionless numbers let us substitute $\sigma_{\rm T}$ from~(\ref{eq:def_gammarad}) and define the cold plasma gyroradius as $\rho_0 = m_e c^2 / e B_0$. We can then rewrite the expression above:
\begin{equation}\label{eq:sim_mult_theory}
    \frac{\eta}{(1+\eta)^{6-4\Gamma}} = 
            \beta_{\rm rec}^3 f_0
            \frac{s_1^2 s_2}{\rho_0^3}
            \frac{\gamma_c^{8-4\Gamma}}{\gamma_{\rm rad}^6\sigma_0^{5-4\Gamma}},
\end{equation}
or, assuming $\eta \gg 1$, 
\begin{equation}\label{eq:sim_mult_theory2}
    \eta = \left(\beta_{\rm rec}^3 f_0\right)^{1/(4\Gamma-5)}
            \left(\frac{s_1^2 s_2}{\rho_0^3}\right)^{1/(4\Gamma-5)}
            \frac{\sigma_0}{\gamma_{\rm rad}^{6/(4\Gamma-5)}\gamma_c^{(4\Gamma-8)/(4\Gamma-5)}}.
\end{equation}
We will compare this estimate with our simulations in section \S\ref{sec:accell} where we describe our results.

In $\gamma$-ray pulsars the power law index of photons is close to $\Gamma\approx 1\text{-}2$ \citep{2013ApJS..208...17A}. To estimate the multiplicity near the light cylinder, we can use the values of $\gamma_c$ and $\gamma_{\rm rad}$ defined in Eq.~(\ref{eq:rad_c_values}) and take $n_0 = \kappa n_{\rm GJ}$, where $\kappa\approx 10^4$ is the multiplicity of the primary cascade near the polar cap, and $n_{\rm GJ}\approx\Omega B_0 / 2\pi c e$. We will also assume that $s_1 = s_2 \approx 0.1 R_{\rm LC}$. We then find, for $\Gamma=2$, that
\begin{equation}\label{eq:obs_mult_theory}
    \eta \approx 180~\left(\frac{B_0}{10^5~\text{G}}\right)^{5/2}\left(\frac{P}{100~\text{ms}}\right)^{3/2}\frac{(s_1^2 s_2)^{1/3}}{0.1~R_{\rm LC}}.
\end{equation}
In particular, for the Crab ($B_0\sim 10^6$ G, $P=33$ ms) we find $\eta\sim 10^4$ for $\Gamma=2$. Thus, we expect pair loading to significantly affect reconnection near the Y-point.

\section{Simulation setup}
\label{sec:setup}

We set up a 2D relativistic particle-in-cell simulation of a reconnecting current sheet using the code {\it TRISTAN-MP} \citep{2005AIPC..801..345S}. The current sheet is initially in Harris equilibrium and is perturbed by either the artificial cooling of the central region or by adding a temporary magnetic loop near the center (see, e.g.,~\citealt{2018MNRAS.473.4840W}). In either case, the evolution does not depend on the way we trigger reconnection, and the later steady state behavior is completely determined by our initial parameters.

Reconnection rapidly develops into the non-linear stage where plasmoids are formed, advecting magnetic field loops and particles out of the box. While in one direction the absorbing boundary conditions allow plasma and fields to flow out (left and right boundaries on Figure~\ref{fig:rec_scheme}), in the other direction the boundaries are being constantly extended while injecting seed primary plasma at rest in the far upstream (for more details see~\citealt{2014ApJ...783L..21S}). This approach eliminates reflections from the boundaries.

The box size, $L$, is chosen so that $L/\sigma\rho_0 \gtrsim 50$, where $\rho_0$ is the gyroradius of a low energy particle in the upstream field, and $\sigma$ is either the initial upstream magnetization parameter, $\sigma_0$, or, in the case of pair production, the effective magnetization, $\sigma_{\rm eff}$, which is much lower due to the mass loading by secondary pairs. After the reconnection starts, we wait until the transient plasmoids are advected out of the box, and the steady state is reached, and then turn on the cooling and pair production. We then wait for a few light-crossing times of the box until a new steady state is reached.

A charged ultra-relativistic particle ($\gamma\gg 1$, $\beta\approx 1$) in electromagnetic field experiences the radiation drag force (see, e.g., \citealt{1975ctf..book.....L})
\begin{equation}
\label{eq:drag}
    \bm{F} = -\frac{2}{3}r_e^2\gamma^2\bm{\beta}\left[\left(\bm{E}+\bm{\beta}\times\bm{B}\right)^2 - \left(\bm{\beta}\cdot\bm{E}\right)^2\right],
\end{equation}
where $r_e$ is the classical radius of the electron, and radiates synchrotron photons. In our simulations we define the effective perpendicular magnetic field,
\begin{equation}
    B_{\rm eff}^2=\left(\bm{E}+\bm{\beta}\times\bm{B}\right)^2 - \left(\bm{\beta}\cdot\bm{E}\right)^2,
\end{equation}
which we will further use to compute the photon energy and the cooling rate. When the synchrotron cooling is enabled, each particle in our simulation probabilistically radiates a photon at the corresponding synchrotron frequency set by the pair threshold parameter, $\gamma_c$,
\begin{equation}
    \label{eq:synch_form}
    \varepsilon_{\rm sync} = m_e c^2\left(\frac{\gamma}{\gamma_c}\right)^2\left(\frac{B_{\rm eff}}{B_0}\right)\left(\frac{\sigma_0}{10^3}\right)^{1/2},
\end{equation}
where $B_0$ is the upstream magnetic field, and the last factor takes into account the magnetic field normalization, i.e., $\varepsilon_{\rm sync}\propto \gamma^2 B_0$. The photon radiation probability is set in such a way that the overall cooling rate, or the energy lost by a particle in a timestep, is consistent with the drag force in formula (\ref{eq:drag}) and is controlled by $\gamma_{\rm rad}$:
\begin{equation}
    \label{eq:synch_rate}
    \frac{\Delta \varepsilon}{\Delta t} \propto \left(\frac{\gamma}{\gamma_{\rm rad}}\right)^2 \left(\frac{B_{\rm eff}}{B_0}\right)^2.
\end{equation}\par
Our simulation tracks photons as regular chargeless and massless particles propagating along straight lines with speed $c$. We do not track photons with very low energies (typically lower than $0.1\text{-}1\%$ of $m_e c^2$), since they do not significantly contribute to pair production (for details see appendix~\ref{sec:appendixB}); they are, however, accounted for in particle cooling.

In each cell, every pair of photons can pair produce with a certain probability. We compute these binary probabilities for each pair of photons in the same cell according to Breit-Wheeler cross section and probabilistically create electron-positron pairs in that cell with momenta consistent with the differential cross section of the process. This algorithm naturally conserves both momentum and energy. Since we loop through all pairs of photons in each cell, this automatically ensures that the mean free path of every photon is inversely proportional to the number density of photons.

Magnetic reconnection in real astrophysical environments is controlled by a combination of the inflowing plasma density and the background magnetic field. Corresponding synchrotron energy is coupled to these parameters through the Planck's constant, $\hbar$, while pair production cross section magnitude is set by the Thomson cross section, $\sigma_{\rm T}$. Particle-in-cell simulations without radiation can be made dimensionless by normalizing simulation parameters to plasma length- and time-scales. Radiation and pair production introduce two extra scales (set by $\hbar$ and two-photon interaction cross section). We thus have four dimensionless parameters in our simulations:
\begin{itemize} 

    \item The cold upstream magnetization parameter, $\sigma_{\rm c} = \sigma_0$, and the radiation-reaction limit, $\gamma_{\rm rad}$, set the background magnetic field and the effective plasma density, determining the available magnetic energy budget per particle and the rate with which particle energy is being transferred to radiation. 

    \item The pair threshold parameter, $\gamma_c$, sets the scale of the Schwinger field, $B_{\rm S}$, with respect to the background magnetic field through relation (\ref{eq:def_gammas}). Given the total energy deposited into radiation (which is set by $\sigma_0$ and $\gamma_{\rm rad}$), this parameter determines how this energy is distributed between photons, defining the Lorentz-factor of the particle that radiates an MeV photon in the upstream field, as defined in equation (\ref{eq:def_gammac}). 

    \item The effective mean free path to pair production is set by the fiducial dimensionless parameter $p_0$. In particular, the probability for two photons to produce $e^-e^+$ pair in one timestep is
    \begin{equation}
        p(\Delta t)=p_0\frac{\sigma_{\gamma\gamma}}{\sigma_{\rm T}},
    \end{equation}
    where $\sigma_{\gamma\gamma}$ is the Breit-Wheeler cross section described in \S\ref{subsec:twophotpp} that depends on momenta and energies of the two photons, and $\sigma_{\rm T}$ is the Thomson cross section. We choose a particular value of $p_0$ to ensure the small optical thickness of the overall system, while still producing enough secondary plasma. In Appendix~\ref{sec:appendixA} we discuss in more detail how varying this parameter affects the regime we are simulating. 
\end{itemize}

In order to accurately sample the photon momentum space in our typical runs, we would require on average $10^3\text{-} 10^5$ photons per cell. To make this problem computationally feasible, we introduce ``weights'' for each photon and a merging routine to combine them. We use the algorithm described by \cite{2015CoPhC.191...65V} that merges all photons in the same momentum bin (and the same cell) to produce two ``heavy'' photons, while conserving energy and momentum without introducing any unnecessary momentum spread. Our momentum bins are uniform in direction and logarithmic in energy. To avoid artificial effects of binning, such as clumping of momenta in a particular direction, we change the orientations and sizes of these bins with times.

The disadvantage of this photon merging process is that it artificially decreases the pair production efficiency, since photons merged into a single ``heavy'' photon can no longer pair produce with each other. This turns out to have a weak effect on the results, since all the merged photons have small angles between their momenta and small relative energies, and so they have a small interaction probability according to Figure~\ref{fig:sigma_pp}. While photon merging is as expensive as pair production in terms of computational time, it is crucial to implement to reduce the memory usage. As a result, we can reduce the number of photons by a factor of $10\text{-}100$, while keeping the overall energy and momentum conserved and not affecting strongly the pair production. 

In addition, we make several compromises in order to reduce both the time and memory consumption.
Synchrotron cooling makes plasmoids effectively compressible, and in the strong cooling regime the centers of plasmoids may get very high overdensities where the local skin depth becomes unresolved:
\begin{equation}
    \frac{c/\omega_{\rm p}}{c/\omega_{\rm p0}} = \langle \gamma \rangle^{1/2} \left(\frac{n_0}{n}\right)^{1/2}.
\end{equation} 
Here, $\langle \gamma \rangle$ is the mean Lorentz-factor of particles, which is typically of the order of $\sigma_0$ in plasmoid centers, $n/n_0$ is the particle overdensity compared to the upstream, $c/\omega_{\rm p}$ and $c/\omega_{\rm p0}$ are the local and upstream skin depths. To resolve the local skin depth with the same number of cells as in the upstream, we need to prevent overdensities larger than $\sim \sigma_0$. To alleviate this issue, we turn off the cooling in the very centers of plasmoids, where the overdensity compared to the upstream is above a certain value (typically few $\sigma_0$). The plasma cools down within the plasmoid until it reaches a particular distance from the center below which the cooling is turned off. The hot central region supports the plasmoid against further contraction; this has no noticeable effect on plasmoid motion. Upstream pair production is also not strongly affected by this effect, since the photons emitted in the centers would not escape the plasmoid due to a high optical depth inside. If the cooling were not turned off in the center, radiated photons would pair produce inside the plasmoid, feeding its interior with additional plasma until the plasmoid center cooled down and stopped emitting pair producing photons.

We also allow photons to escape into upstream from within the dense regions of plasmoids before they pair produce. This results in the lack of secondary plasma in the plasmoid interiors which, however, does not affect the overall physics, since most of this plasma would be carried away with the plasmoids,  unable to escape upstream. 

\section{Results}
\label{sec:results}

\begin{figure*}[tb]
    \centering
    \centerline{\includegraphics[width=2\columnwidth]{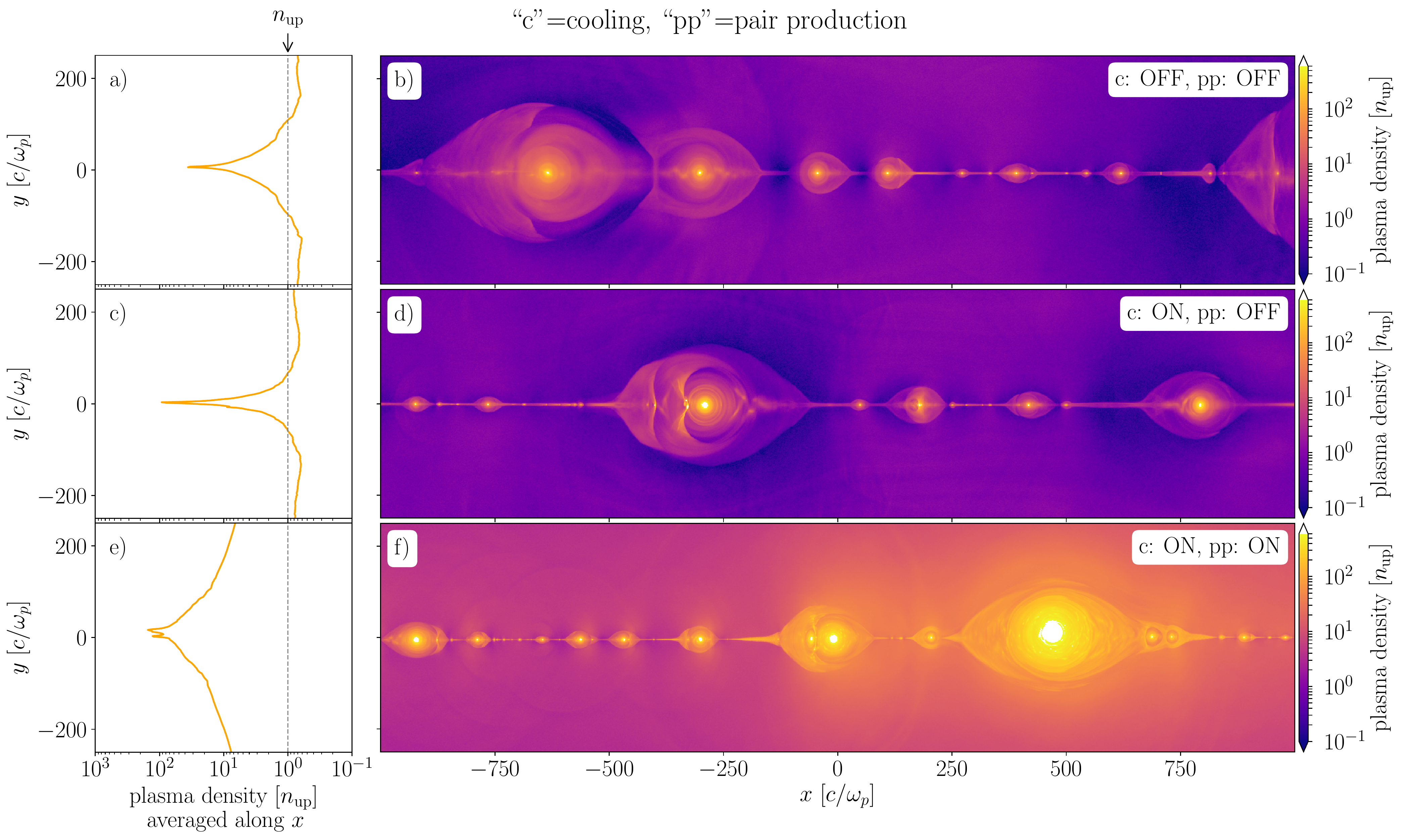}}
    \caption{Snapshots of plasma density from three different simulations with $\sigma_0=5000$: a-b) without synchrotron cooling or pair production; c-d) with cooling ($\gamma_{\rm rad}=1000$, $\gamma_c=50$) but without pair production; e-f) with both cooling and pair production (additionally $p_0=10^{-5}$). Three plots on the left (panels a, c, e) represent the averaged densities along the current sheet ($x$ axis) plotted vs $y$ axis. Note that all three colorbars as well as the horizontal axes in panels (a, c, e) have the same scale for convenience. When cooling is enabled, plasmoids become effectively compressive, so the typical plasmoid sizes and central overdensities are larger in panel (c) compared to panel (a), where there is no cooling. With pair production enabled (panels e, f) the secondary plasma mass loads both the far upstream and plasmoids, making them larger and heavier.}
    \label{fig:snapshot_dens}
\end{figure*}

\begin{figure*}[tb]
    \centering
    \centerline{\includegraphics[width=1.7\columnwidth]{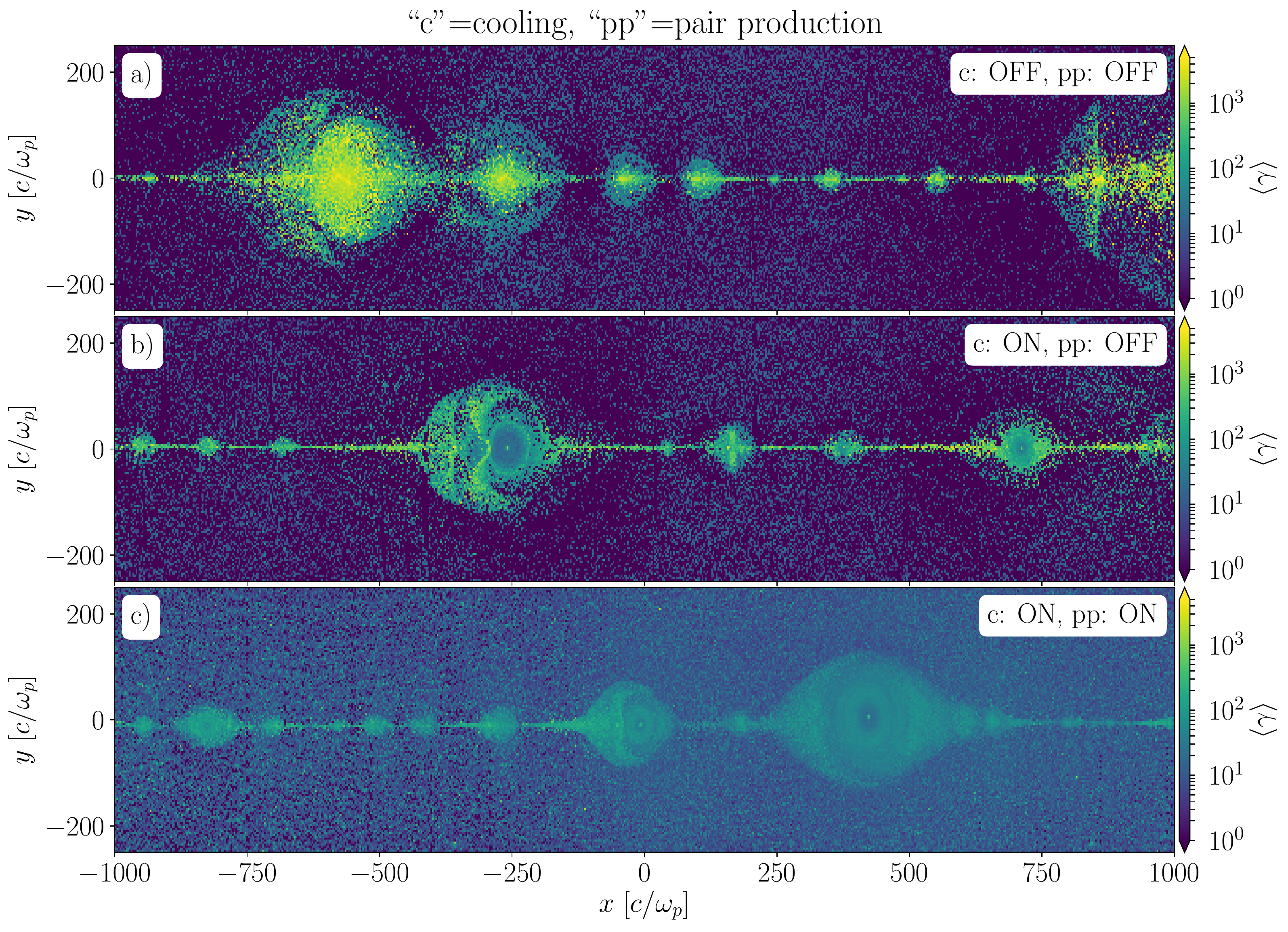}}
    \caption{Snapshots of the mean particle Lorentz factor from the same three simulations as in Figure~\ref{fig:snapshot_dens}. In panel (a) where we present the fiducial run without cooling or pair production, most of the high energy plasma with $\gamma\sim \sigma_0$ is contained within the plasmoids, providing pressure support to resist contraction. When synchrotron radiation is enabled, shown in (b), particles within the plasmoids are efficiently cooled to energies $<\gamma_{\rm rad}$, the pressure support is weaker, and the plasmoids are denser. In this case most of the high energy plasma is contained around the primary current layer, where the magnetic field is weak and the cooling is inefficient. Panel (c) shows the run with pair production, in which case the upstream is hotter, since the newly born secondary pairs inherit the power-law energy distribution from their parent photons.}
    \label{fig:snapshot_meang}
\end{figure*}


In this section we will discuss our main simulation results. We first compare how synchrotron cooling and pair production individually affect the overall dynamics of reconnection and the formation and evolution of plasmoids. We then turn to particle acceleration and the emerging particle and photon spectra, again considering the effects of cooling and pair production separately. In most of our simulations we take the upstream magnetization to be $\sigma_0\sim 10^3\text{-}10^4$, in order to reach the desired high secondary plasma multiplicities. Cooling parameters $\gamma_c$ and $\gamma_{\rm rad}$ are typically a few times $10^2\text{-}10^3$, to ensure that the number of high energy photons is large enough and the cooling is in the strong regime. The cross section magnitude, $p_0=10^{-5}$, is chosen so that enough pairs are produced (multiplicity $\gg 1$), but at the same time the optical depth to pair production is small across the box for all photons ($\tau_{\gamma\gamma}<1$, see details in appendix~\ref{sec:appendixA}).

\subsection{Upstream inflow and plasmoids in the steady state}

In Figure~\ref{fig:snapshot_dens} we show a snapshot of the plasma density (panels b,d,f) and the plot of the density averaged along the current sheet (panels a,c,e) for three different cases. For all the runs we use 3 cells per cold skin depth, $c/\omega_{\rm p0}$, initial plasma has 5 particles per cell, and the box size is  $L=7500~\text{cells}=2500~c/\omega_{\rm p0}$. Our results are unchanged for different values of skin depth ($3$ to $10$ cells), and the number of particles per cell ($2$ to $10$). As a fiducial case, in Fig.~\ref{fig:snapshot_dens}a,b we show relativistic reconnection ($\sigma_0=5000$) without either synchrotron cooling or pair production. In Fig.~\ref{fig:snapshot_dens}c,d we show the result with synchrotron cooling turned on ($\gamma_{\rm rad}=1000$, $\gamma_c=50$), but without pair production, and finally Fig.~\ref{fig:snapshot_dens}e,f show the run with pair production. Also, in Fig.~\ref{fig:snapshot_meang} we show the same snapshots as in Fig.~\ref{fig:snapshot_dens}, but plotting the mean particle Lorentz-factor in each cell to indicate the temperature of the plasma. For ease of comparison all the corresponding colorbar scales are the same.

In the fiducial run (Figs.~\ref{fig:snapshot_dens}a,b, and \ref{fig:snapshot_meang}a), dense plasmoids contain most of the high energy plasma particles (with $\gamma\sim \sigma_0$, see yellow regions in Fig.~\ref{fig:snapshot_meang}a), accelerated in x-points and mergers. This hot plasma provides the pressure support for plasmoids against contraction due to the Lorentz force, $\bm{j}\times\bm{B}$. Interiors of plasmoids typically have higher magnetic fields and densities compared to both the current sheet and the plasmoid peripheries. In the presence of cooling (Figs.~\ref{fig:snapshot_dens}c,d, and \ref{fig:snapshot_meang}b), the high energy particles within the plasmoids will be efficiently cooled to energies $<\gamma_{\rm rad}$. We see in Fig.~\ref{fig:snapshot_meang}b that the interiors of plasmoids have smaller average $\gamma$-factors compared to Fig.~\ref{fig:snapshot_meang}a. This removal of pressure support makes plasmoids compressive: they are now slightly smaller in size and are more dense and concentrated toward their centers (see Fig.~\ref{fig:snapshot_dens}c).

When pair production is enabled (Figs.~\ref{fig:snapshot_dens}e,f and \ref{fig:snapshot_meang}c), photons radiated in the current sheet and plasmoids escape to the upstream and pair produce with each other, additionally loading the system with secondary plasma (see Fig.~\ref{fig:snapshot_dens}f: both upstream and plasmoids are significantly denser compared to Fig.~\ref{fig:snapshot_dens}b,d). These upstream pairs inherit the power law distribution function from the parent photons and are hot, as opposed to the cold inflowing primary plasma injected at the boundaries. This hot upstream can be seen as higher average Lorentz factor plasma in the upstream region in Fig.~\ref{fig:snapshot_meang}c.

\begin{figure*}[htb]
    \centering
    \centerline{\includegraphics[width=2.2\columnwidth]{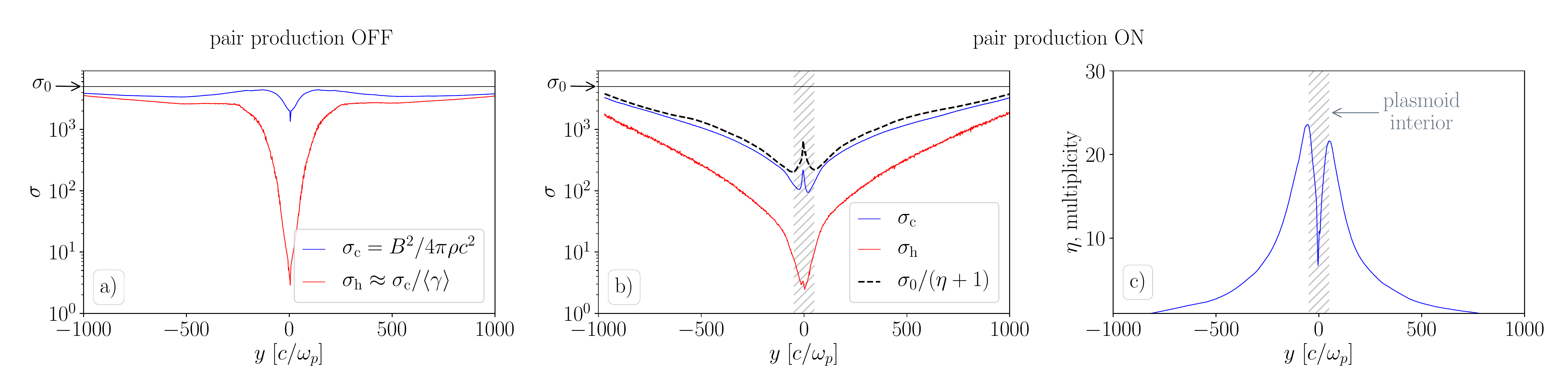}}
    \caption{Cold magnetization parameter, $\sigma_{\rm c}=B^2/4\pi\rho c^2$, and hot magnetization parameter, $\sigma_{\rm h}=B^2/4\pi\langle\gamma\rangle\rho c^2$, averaged along the current sheet and over time for the runs without (a) and with pair production (b). Panel (c) shows the average multiplicity (density of secondary pairs divided by the density of primary pairs). Shaded regions show the interior of plasmoids where the multiplicity is suppressed artificially. Two solid black lines show the level of the upstream magnetization $\sigma_{\rm c}$, which, when divided by multiplicity, gives the effective magnetization (black dashed line). In contrast to the case without pair production (a), when secondary plasma is produced the magnetization is reduced by a factor of multiplicity (b). Also, since the newly born pairs are hot, $\sigma_{\rm h}$ is also reduced in the upstream.}
    \label{fig:sigmas_cut}
\end{figure*}


Pair production also decreases the effective magnetization. To demonstrate this effect, in Figure~\ref{fig:sigmas_cut} we plot the cold, $\sigma_{\rm c}\equiv B^2/4\pi\rho c^2$, and hot, $\sigma_{\rm h}=\sigma_{\rm c}/\langle\gamma\rangle$, magnetization parameters, as well as the multiplicity of the secondary plasma, averaged along the $x$-axis and plotted against the $y$-axis.  Fig.~\ref{fig:sigmas_cut}a is the fiducial run without pair production or cooling, while Fig.~\ref{fig:sigmas_cut}b,c are for the run with pair production. 

In Fig.~\ref{fig:sigmas_cut}a we see that the cold magnetization parameter is very close to the far upstream value, $\sigma_0$, shown with a black solid line. On the other hand, pair production suppresses this number (Fig.~\ref{fig:sigmas_cut}b) by a factor of $\eta +1$, where $\eta$ is the effective multiplicity, also computed {\it in situ}\footnote{The local multiplicity here is defined as the density of secondary pairs produced via two-photon pair production divided by the density of primary pairs originally injected from the boundaries.} (see Fig.~\ref{fig:sigmas_cut}c). Note also that the hot magnetization, $\sigma_{\rm h}$ (red line), in the case with pair production, differs strongly from $\sigma_{\rm c}$ (blue line) even in the upstream. This underscores the fact that the upstream plasma, which is dominated by secondary pairs, is hot.

\begin{figure*}[tb]
    \centering
    \centerline{\includegraphics[width=2\columnwidth]{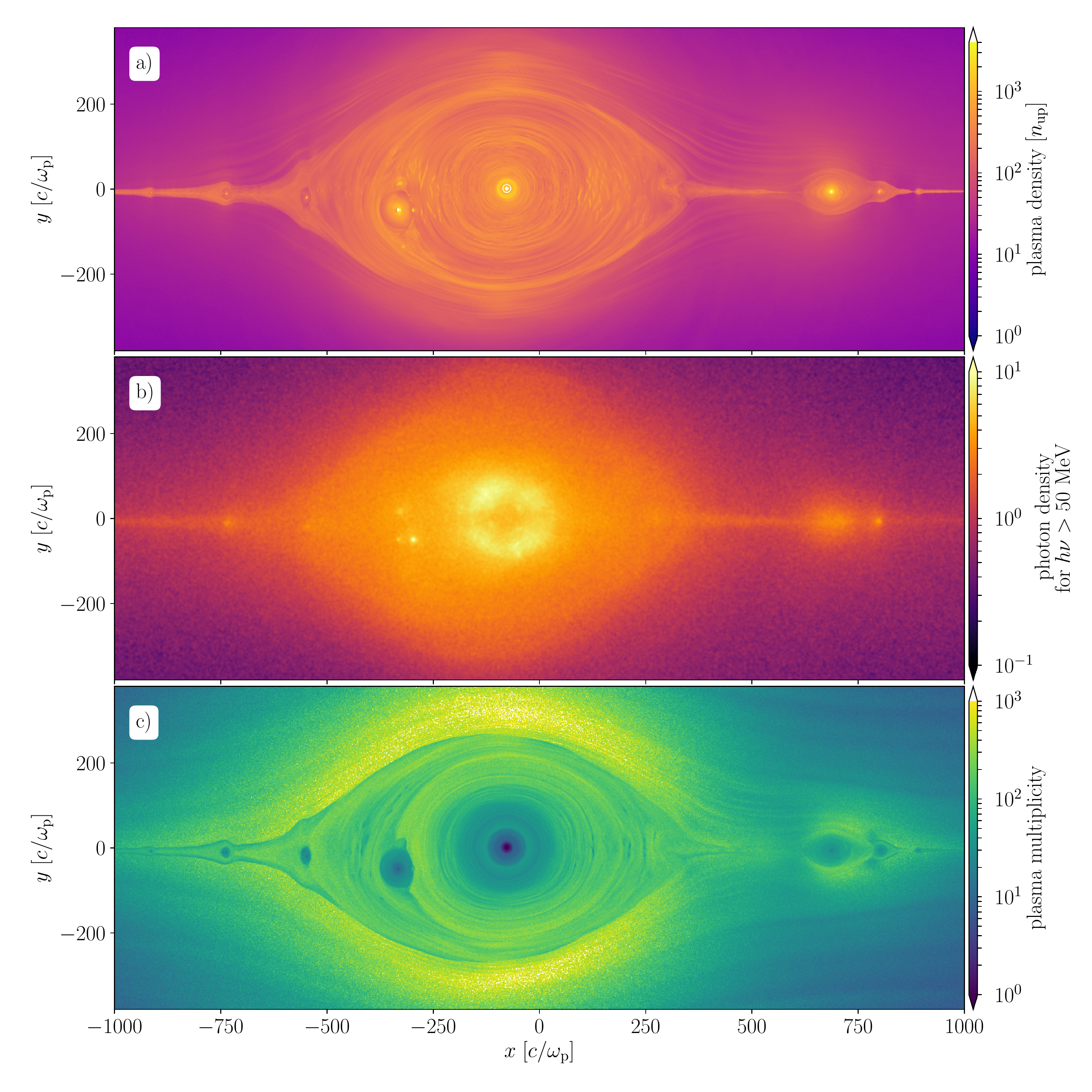}}
    \caption{Snapshot from a simulation of the reconnection with the cooling and pair production with $\sigma_0=10^4$, $\gamma_c=50$, $\gamma_{\rm rad}=1000$, $p_0=10^{-5}$. The box extends further in the vertical direction, in this figure we focus on the large plasmoid. Plots show (a) the total plasma density normalized by the initial upstream density, (b) the density of photons with $h\nu>50$ MeV, (c) the local plasma multiplicity, $\eta$, defined as the density of secondary plasma divided by the density of the injected plasma. Interiors of plasmoids are low in multiplicity since we allow photons to escape from them without having a chance to pair produce. Maximum density in the top plot (in white) indicates the region, where we turn off cooling to prevent plasmoids from getting too dense and making the local skin depth unresolved. When the pair production is enabled the upstream is abundant with secondary plasma and also plasmoids are typically larger and slower, since they rapidly expand via pair production. In this snapshot the massive plasmoid in the middle is undergoing mergers with smaller plasmoids, one of which (at $x=-300$, $y=-50$) has not yet been fully absorbed. Most of the photons are emitted within plasmoids close to their centers, where both the magnetic field and density are high. Lack of high energy photons in the very middle of the large plasmoid is due to the fact that plasma in that region is cold. The full movie of this simulation is available \href{\linkVidOne}{online} (direct link: \url{\linkVidOne}).}
    \label{fig:snapshot}
\end{figure*}

Pair production also has a strong impact on the overall dynamics and evolution of plasmoids. Figure~\ref{fig:snapshot} shows a typical snapshot from one of our pair producing simulations with higher $\sigma_0$. In this figure we show the plasma density (Fig.~\ref{fig:snapshot}a), the density of high energy photons with $h\nu>50$ MeV (Fig.~\ref{fig:snapshot}b) and the local multiplicity (Fig.~\ref{fig:snapshot}c).
Without pair production the plasmoids can grow only through the slow accretion of plasma from upstream and through plasmoid mergers. When pair production is enabled, plasmoids radiate high energy photons (Fig.~\ref{fig:snapshot}b), some of which can pair produce with the low energy photons near the plasmoid peripheries (see high multiplicity regions in Fig.~\ref{fig:snapshot}c). This additional mass loading with hot secondary plasma produces more radiation that pair produces further, and the plasmoid continues to expand as it is being advected out of the simulation box. 

In the real reconnection this plasmoid growth will cease once the emitted photons can no longer pair produce near plasmoid peripheries. This can happen either because the optical depth across the plasmoid is large and the photons never escape from it, or because the plasmoid cools down, and the radiated photons have energies that are too low. In our simulations we allow these high energy photons to escape from the plasmoids before they pair produce. 

In addition, the mergers constantly eject new high energy photons from freshly accelerated particles. These photons can then pair produce outside the current sheet, providing even more secondary plasma to the plasmoids. Because of this, plasmoids are typically larger and slower for higher multiplicities. When the sizes of plasmoids become comparable to the simulation box size, they effectively slow down the reconnection rate and in some situations can nearly stop it. After plasmoids leave the simulation box, the reconnection continues with the regular rate at $\beta_{\rm in}\approx 0.1\text{-}0.2$. The movie of our simulation from Fig.~\ref{fig:snapshot} that shows a large plasmoid and explosive photon ejections during plasmoid mergers is available \href{\linkVidOne}{online}\footnote{Direct link: \url{\linkVidOne}.}.


\subsection{Particle acceleration and the emerging radiation spectra}
\label{sec:accell}

\begin{figure*}[tb]
    \centering
    \centerline{\includegraphics[width=2.2\columnwidth]{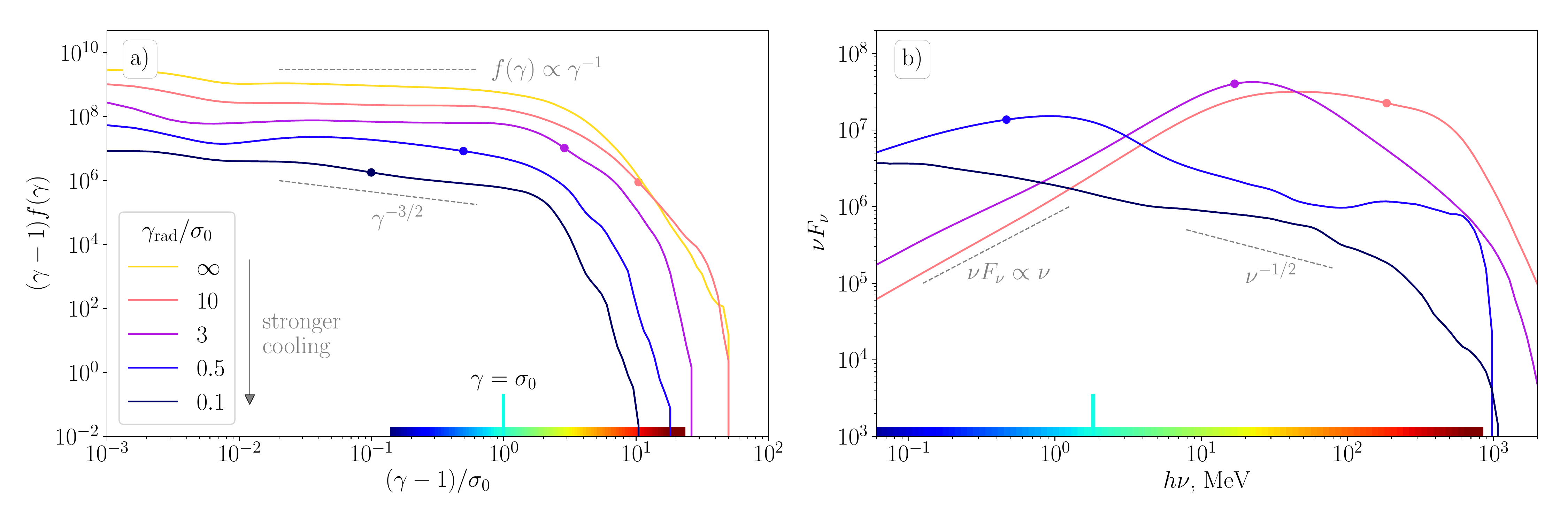}}
    \caption{Particle and photon spectra for the $\sigma_0=200$ simulations without pair production for different cooling regimes in late-time steady state. The cooling decreases, i.e., $\gamma_{\rm rad}$ increases, from dark blue to yellow; $\gamma_{\rm rad}=\infty$ is the case without cooling (i.e., no photons). Two colorbars below put particle energies and synchrotron photon energies into correspondence, given a fixed background magnetic field, which in general varies depending on particle position. Particle spectra are shifted vertically for illustration purposes; circles indicate $\gamma_{\rm rad}$ for each case as well as the corresponding synchrotron photon energy in the upstream magnetic field. Cooling typically conserves the $\gamma^{-1}\text{-}\gamma^{-3/2}$ slope of the particle energy distribution function. Peaks in synchrotron photon spectra for each case corresponding to $\gamma_{\rm rad}$ in particles; high energy tails with $\nu F_{\nu}\propto \nu^{-1/2}$ are formed during transient mergers, when particles accelerated in secondary current sheets are captured by plasmoids and rapidly cooled.}
    \label{fig:no_pair_runs}
\end{figure*}

We now focus on how cooling and pair production affect particle acceleration in reconnection, and how the resulting particle and photon spectra are formed. To distinguish the effects of cooling and pair production we first analyze the process without pair production. We test how the cooling regime affects particle and radiated photon spectra (cf., e.g., \citealt{2016ApJ...833..155K, 2018JPlPh..84c7501N}).  

The cooling regime is characterized by the ratio $\gamma_{\rm rad}/\sigma_0$, which effectively sets the cooling time compared to the acceleration time $\sim (t_{\rm cool}/t_{\rm acc})^{1/2}$. If this ratio is $\gg 1$, the cooling is inefficient, i.e., the cooling time is much longer than the acceleration time, and vice versa. Note, however, that this ratio is defined for the upstream magnetic field, so the actual cooling time of a particle will depend on the local magnetic field.

We initialize our simulations with the magnetization parameter $\sigma_0=200$ and cold plasma in the upstream. We fix $\gamma_c=50$ and vary $\gamma_{\rm rad}$ to study the different regimes. The results of our runs are shown in Figure~\ref{fig:no_pair_runs}, where we plot the particle and photon spectra for the runs in different cooling regimes: from no cooling (yellow line) to strong cooling (dark blue line). The horizontal colorbar puts into correspondence the particle energies and the energies of synchrotron photons that would have been radiated by these particles in the upstream magnetic field; circles on the lines mark $\gamma_{\rm rad}$ and the corresponding synchrotron photon energy.

When the cooling is weak, $\gamma_{\rm rad}/\sigma_0\gg 1$, particles form a power-law, $f(\gamma) \propto \gamma^{-1}$, similar to what is observed in uncooled simulations for high $\sigma_0$ (cf., e.g., \citealt{2014ApJ...783L..21S}). The strong cooling cases have slightly steeper spectra (close to $f(\gamma)\propto \gamma^{-3/2}$). For all values of $\gamma_{\rm rad}$ the power-law slope is followed by a steep decay at higher energies (close to a few times $\sigma_0$, see Fig.~\ref{fig:no_pair_runs}a).

The overall picture is then as follows: particles are accelerated in primary and secondary current sheets up to a few times $\sigma_0 m_e c^2$. The magnetic field in these current sheets is weak, and the cooling time, $t_{\rm cool}$, is much longer than the residence time in the x-points. This means that the cooling is inefficient close to the current sheets, and the particles remain hot until they are captured by the plasmoids, where they cool down to energies $<\gamma_{\rm rad}$. Since the acceleration and cooling happen in different regions, we do not expect a different power-law slope or any pronounced power-law break in the particle distribution function close to $\gamma_{\rm rad}$. The cutoff energy in particles is dictated by the acceleration in x-points, and is typically a few times $\sigma_0$. Note that there is no growth of this cutoff over time due to plasmoid compression as found in the uncooled simulations of \cite{2018arXiv180800966P}, because particles within the plasmoids are efficiently cooled to $\gamma_{\rm rad}$ in our runs, unless the cooling is too weak $\gamma_{\rm rad}/\sigma_0\gg 1$.

The corresponding time-averaged photon spectra are shown in Fig.~\ref{fig:no_pair_runs}b. A standard result in the theory of synchrotron radiation is that particles constantly injected with a power-law energy distribution $f(\gamma)\propto \gamma^{-p}$ form a synchrotron spectrum described as $\nu F_{\nu}\propto \nu^{-(p-3)/2}$ \citep{1979rpa..book.....R}. In our simulations, the plasma that resides within the plasmoids ($\gamma < \gamma_{\rm rad}$)  and forms a power-law spectrum with $p\approx 1$, radiates $\nu F_{\nu}\propto \nu$ synchrotron spectrum (see photon spectra before the peak in Fig.~\ref{fig:no_pair_runs}b). The peaks in photon spectra correspond to particles with the energy $\gamma_{\rm rad}$ radiating in magnetic fields of value close to the upstream field.

Photons with energies beyond the peak frequency are radiated by the particles with energies up to a few times $\sigma_0$. These particles enter the plasmoids after being accelerated in x-points and rapidly lose their energy, being exposed to a large perpendicular magnetic field, until they cool down to $\gamma\lesssim \gamma_{\rm rad}$. As a result, the photon spectrum at high energies is highly fluctuating in time, since there is no steady population of high energy particles ($\gamma > \gamma_{\rm rad}$) in a large magnetic field. The majority of these high energy photons, especially for the strong cooling regime ($\gamma_{\rm rad}/\sigma_0\ll 1$), are radiated during violent events such as plasmoid mergers. Averaged over time,  these transients form an extended power law tail (close to $\nu^{-1/2}$). We also present these results in the form of movies to better illustrate how the transients work both in \href{\linkVidThree}{strong} and \href{\linkVidTwo}{weak} cooling regimes\footnote{Direct links: \url{\linkVidThree} and \url{\linkVidTwo}.}.

Further, we discuss the results of our runs with pair production. For all the runs below we chose the parameters to be $\sigma_0=10^3\text~\text{to}~4\times10^4$, $\gamma_{\rm rad} = 10^3$, $\gamma_c=50$ and $p_0=10^{-5}$. These values are chosen to satisfy the relation $\gamma_c \ll \gamma_{\rm rad} \lesssim \sigma_{\rm eff}$ described in \S\ref{subsec:pcoolandphot}. We also want the synchrotron peak to be roughly at $0.1\text{-}1$ GeV, which fixes $\gamma_c$ and $\gamma_{\rm rad}$. 

\begin{figure*}[htb]
    \centering
    \centerline{\includegraphics[width=2.2\columnwidth]{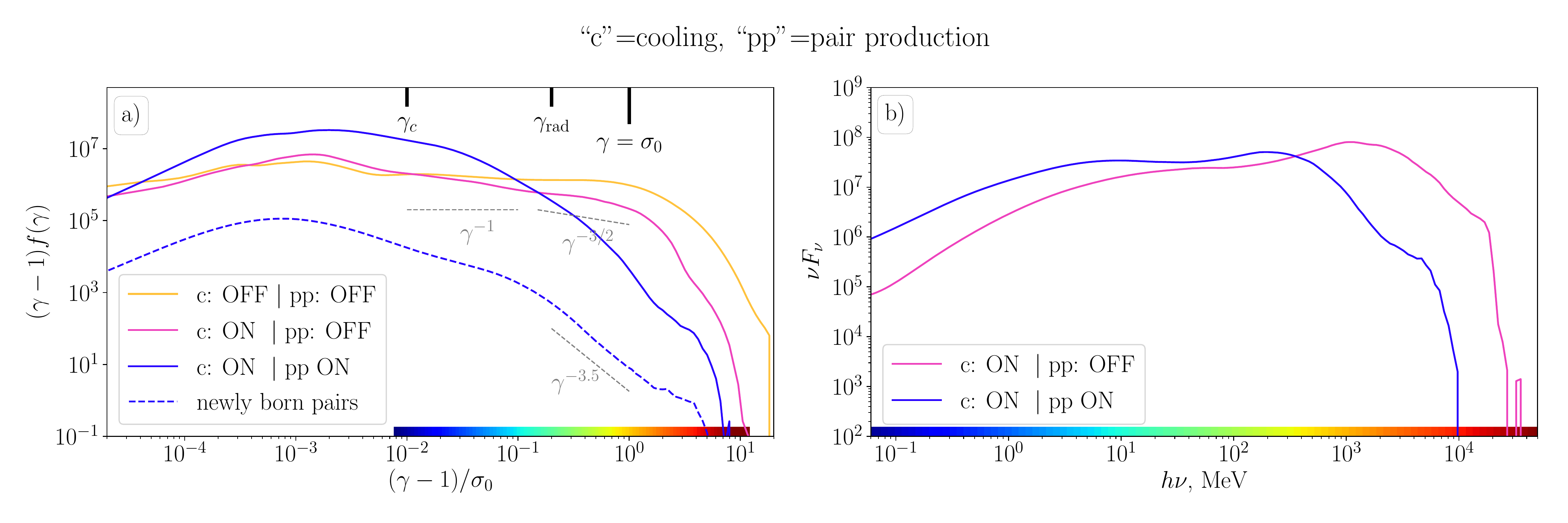}}
    \caption{Particle (a) and photon (b) spectra for the $\sigma_0=5000$, $\gamma_{\rm rad} = 1000$, $\gamma_c=50$ simulations without cooling, with cooling but without pair production, and with both cooling and pair production enabled in late-time steady state. Two colorbars below put particle energies and synchrotron photon energies into correspondence, given a fixed background magnetic field. Dashed line is the spectrum of newly born secondary pairs in the run with pair production. Cooling does not strongly affect the slope, whereas when pair production is enabled the effective magnetization is dropped, and the upstream is no longer cold (see the newly born pairs). This causes the shift of the peak in photon spectrum (corresponding to the effective magnetization, $\sigma_{\rm eff}$).}
    \label{fig:pair_runs}
\end{figure*}

We first focus on a single value of $\sigma_0=5000$ to see how enabling pair production changes the shape of the particle and photon spectra. In Figure~\ref{fig:pair_runs} we present the energy distribution of particles (Fig.~\ref{fig:pair_runs}a) and photons (Fig.~\ref{fig:pair_runs}b) for three runs: without cooling or pair production (yellow line), with cooling but without pair production (pink line), and with both cooling and pair production (blue line).

Cooling without pair production does not strongly affect the particle spectrum of $f(\gamma)\propto \gamma^{-1}\text{-}\gamma^{-3/2}$, in agreement with what we found earlier, only slightly decreasing the energy cutoff (see yellow and pink lines in Fig.~\ref{fig:pair_runs}a). On the other hand, when pair production is on, the spectrum is no longer flat and there is no clear cutoff energy. Instead, particles form a wide power-law distribution, ranging from $\gamma^{-2}$ to $\gamma^{-4}$.

There are two effects causing this. First, since the layer is being mass loaded with secondary plasma, the effective magnetization parameter, $\sigma_{\rm eff}$, decreases compared to the upstream magnetization, $\sigma_0$. This means that the reconnection has less magnetic energy per particle and the resulting spectrum shifts to lower energies. At the same time, the newly born secondary pairs (dashed blue line in Fig.~\ref{fig:pair_runs}a) that have not yet been accelerated in the electric field at the x-point, form a steep power law with $f(\gamma)\propto \gamma^{-3}\text{-}\gamma^{-4}$ up to the highest energies. This means that, unlike two other cases where the current sheet was being fed by an initially cold plasma, in this case the upstream secondary plasma is already hot, due to a wide distribution in energies of parent photons that produced these pairs.

This behavior is also reflected in the photon spectrum (Fig.~\ref{fig:pair_runs}b), where the peak is shifted due to the reduced magnetization. The peak in photons corresponds to $\gamma\approx \sigma$ in particles, where in the case of pair production the value of $\sigma$ is reduced with respect to $\sigma_0$ by a factor of multiplicity. In this particular run, presented in Fig.~\ref{fig:pair_runs}, the multiplicity is of the order of a few, and thus the effective magnetization is few times less than $\sigma_0$. The resulting peak frequency is proportional to the particle energy squared, and is, therefore, smaller by roughly an order of magnitude.

\begin{figure*}[htb]
    \centering
    \centerline{\includegraphics[width=2.2\columnwidth]{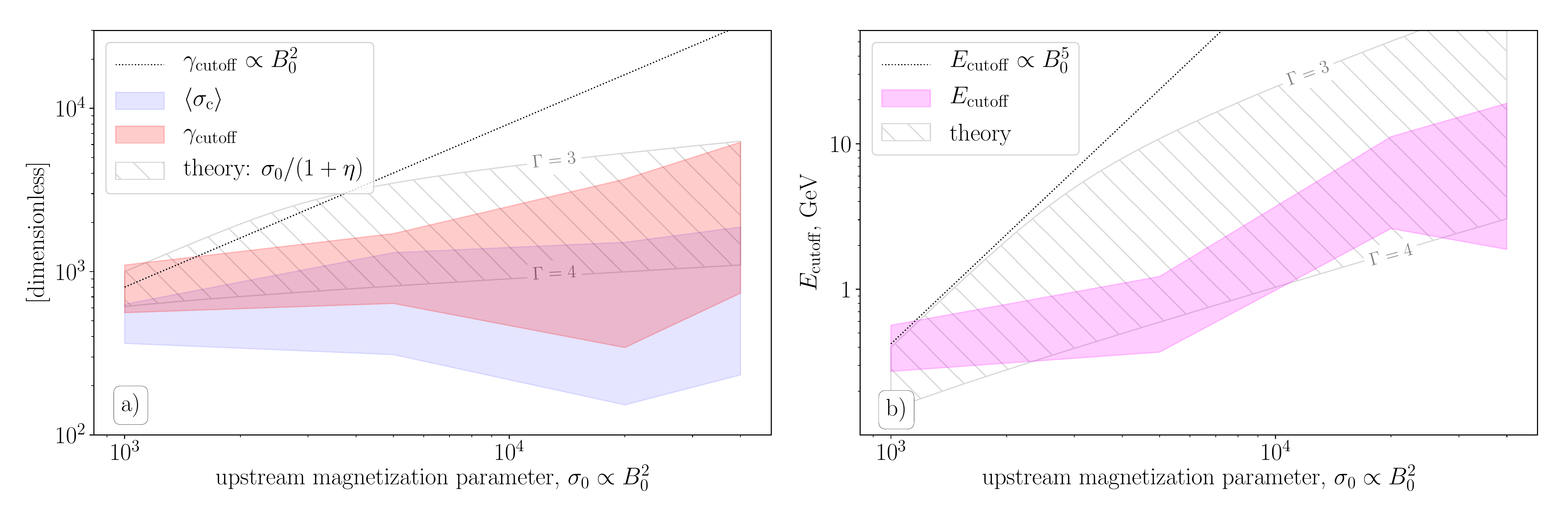}}
    \caption{Box-averaged quantities from four of our simulations: (a) magnetization $\langle\sigma_{\rm c}\rangle$ (blue) and particle spectra cutoff $\gamma_{\rm cutoff}$ (red); (b) photon spectral cutoff $E_{\rm cutoff}$ (magenta), all plotted against the initial magnetization parameter, $\sigma_0$, which is proportional to the background magnetic field squared, $B_0^2$. Widths of the colored regions represent the uncertainties in determining these parameters from simulations. Dotted black lines are simple scalings without pair production feedback, and hatched regions are theoretical predictions from Sec.~\ref{sec:theory} with photon indexes $\Gamma=3\text{-}4$.}
    \label{fig:cutoffs_vs_sigma}
\end{figure*}

Finally, in Figure~\ref{fig:cutoffs_vs_sigma} we show how the particle cutoff, box averaged $\sigma_{\rm c}$ (defined similar to Fig.~\ref{fig:sigmas_cut}b) and spectral cutoff for photons depend on $\sigma_0$, i.e., the background magnetic field $B_0$ (since $\sigma_0\propto B_0^2$ for fixed density). The average cold magnetization, $\langle \sigma_{\rm c}\rangle$ (blue line in Fig.~\ref{fig:cutoffs_vs_sigma}a), is found by averaging $B^2/4\pi\rho c^2$ over a region around the current sheet where most of the pair production takes place. Different sizes of this region give different values, which we represent as errorbars. To determine the cutoffs we define a model-independent measure of how far our spectrum extends 
\begin{equation}
    \gamma_{\rm cutoff}\sim \frac{\int\mathrm{d}\gamma f(\gamma)\gamma^\alpha}{\int\mathrm{d}\gamma f(\gamma)\gamma^{\alpha-1}},
\end{equation}
and the same for photons (similar to what was done by~\citealt{2015ApJ...809...55B}). This gives a rough estimate of where the spectral break (or the cutoff) is. Power $\alpha$ is empirically chosen to be between $2$ and $3$: varying $\alpha$ in that range gives different cutoff energies which we present as errorbars (as a rough estimate of how wide and uncertain the cutoff is).

We compare our findings with the simple predictions without pair production feedback: 
\begin{equation}
    \gamma_{\rm cuttoff}\propto \sigma_0\propto B_0^2,~~~E_{\rm cutoff}\propto \gamma_{\rm cuttoff}^2 B_0\propto B_0^5,
\end{equation}
and also overplot the predicted effective magnetization, $\sigma=\sigma_0 / (1+\eta)$, and the corresponding synchrotron cutoff energies (hatched regions in both plots). For that, we estimate the steady state multiplicity (see equation~\ref{eq:sim_mult_theory}) from the model described in Sec.~\ref{sec:theory} with different values of the photon power law index $\Gamma$.

First, note that both the cutoff energy and the magnetization are reduced compared to simple predictions without pair production feedback. This means that not only is the maximum energy of particles lower with pair production, but there is an effective upstream magnetization, $\sigma_{\rm eff}$, that is decreased from the initial value $\sigma_0$ by a factor of multiplicity of secondary plasma. 

From the runs with different values of $\sigma_0$ we find that the particle cutoff, $\gamma_{\rm cutoff}$, grows roughly as $B_0^{0.2\text{-}0.5}$ with the background magnetic field, and the photon spectral cutoff energy, $E_{\rm cutoff}$, grows as $B_0^{1.2\text{-}1.8}$. These two scalings are consistent with each other, since $E_{\rm cutoff}\propto \gamma_{\rm cutoff}^2 B_0$. We can compare these results with simple predictions described by equation (\ref{eq:sim_mult_theory}). In our simulations the distribution function of high energy photons has a power law index $\Gamma\approx 3\text{-}4$ (corresponding to spectral index $\nu F_{\nu}\propto \nu^{-1}\text{-}\nu^{-2}$). To zeroth order, scale lengths $s_1$ and $s_2$ from  (\ref{eq:sim_mult_theory}) can be taken to be equal to the size of the simulation box $L$. In these simulations we have $L / \sigma \rho_0 \approx 70 (\sigma_0 / 10^3)^{-1/2}$, since the box has a fixed number of cells, but the gyroradius scales with the upstream magnetization. With $s_1 = s_2 = L$ and $\eta \gg 1$, we get $\eta \propto \sigma_0^{0.8}$, meaning that the effective magnetization scales weakly with $\sigma_0$: $\sigma = \sigma_0 / (1+\eta) \propto \sigma_0^{0.2}\propto B_0^{0.4}$ and $E_{\rm cutoff}\propto B_0^{1.8}$.
Taking $f_0\approx 0.1\text{-}0.01$, $\beta_{\rm rec}\approx 0.1$, $\gamma_c = 50$, $\gamma_{\rm rad} = 1000$, and $\Gamma=3\text{-}4$, we find that the steady state multiplicity for $\sigma_0=10^4$ is $\eta\approx 2\text{-}10$, while for $\sigma_0 = 4\times 10^4$ it is $\eta\approx 10\text{-}50$, which roughly agrees with what we see. However, as our simplified model does not account for time variability, we underestimate the multiplicity by a factor of a few, which results in a slightly higher effective magnetization (compared to $\langle\sigma_{\rm c}\rangle$). 

One should also note that in these simulations we only varied the background magnetic field, while keeping the upstream plasma density constant. In the case of pulsars the density of plasma inflowing from the inner magnetosphere to the Y-point also depends on the magnetic field (since the overall plasma density is normalized by Goldreich-Julian density $n_{\rm GJ}=\bm{\Omega}\cdot\bm{B}/2\pi ce$). This will further make the dependence of cutoff energy on the magnetic field weaker.
On the other hand, if the pair production in the outer magnetosphere were not mass loading the current layer, the particle energy cutoff would grow as
\begin{equation}
    \gamma_{\rm cutoff}\propto B_0^2 / n_{\rm LC} \propto B_0,
\end{equation} 
where $n_{\rm LC}\propto n_{\rm GJ}\propto B_0$ is the inflowing density to the current sheet at the light cylinder, which is proportional to the Goldreich-Julian density and thus to the magnetic field strength; and the cutoff energy for photons would grow as $B_0^3$. 

\section{Discussion}
\label{sec:discussion}

In this paper we discussed the results of two dimensional particle-in-cell simulations of radiative relativistic reconnection with pair production. We included both synchrotron cooling and two-photon pair production self-consistently by tracking all the radiated photons as separate particles and colliding them with each other. We then separately studied the effects of both cooling and pair production on particle acceleration and the emission signatures of reconnection. Our main findings are summarized below.

\subsection{Synchrotron cooling} 

\begin{itemize}[leftmargin=*,wide = 0pt]
  \item Particles with energies $\gamma>\gamma_{\rm rad}$ exist near the peripheries of plasmoids or close to the primary or secondary current layers, where cooling is inefficient. Bulk of the particles within the plasmoids have energies $<\gamma_{\rm rad}$.
  
  While in the case of non-radiative reconnection large plasmoids contain the bulk of the high energy particles (with energies up to a few times $\sigma_{\rm c}m_e c^2$), when synchrotron cooling is enabled these particles can no longer maintain their energies within the plasmoids due to the strong magnetic field. Thus, the particles in plasmoids are cooled to energies $\gamma < \gamma_{\rm rad}$, while the rest of the high energy plasma, $\gamma > \gamma_{\rm rad}$, exists either around the primary and secondary current sheets or in the vicinities of plasmoids, where the magnetic field is weak, and the cooling is inefficient. 
\end{itemize}

\begin{itemize}[leftmargin=*,wide = 0pt]
  \item Cooling removes the pressure support for plasmoids against contraction, and plasmoids become effectively compressive with typically smaller sizes and larger overdensities in the centers.
\end{itemize}

\begin{itemize}[leftmargin=*,wide = 0pt]
  \item Weak cooling, $\gamma_{\rm rad}/\sigma_{\rm c}\gg 1$, preserves the hard power-law in particle energy distribution function $f(\gamma)\propto \gamma^{-1}$; in strong cooling regime, $\gamma_{\rm rad}/\sigma_{\rm c}\ll 1$ the slope steepens towards $f(\gamma)\propto \gamma^{-3/2}$.
  
  Since the acceleration and cooling of the particles take place in different locations (x-points and plasmoids), there is no cooling break near $\gamma_{\rm rad}$ and the power-law slope of the particle spectrum is generally unaffected.
\end{itemize}

\begin{itemize}[leftmargin=*,wide = 0pt]
    \item The high energy cutoff in particle distribution is only slightly shifted towards the lower energies (still being a few times $\sigma_{\rm c}$). Because of that, the corresponding cutoff in photon spectrum is only marginally affected by cooling.
    
    This maximum energy is roughly unaffected, because the x-points are still able to accelerate particles effectively up to a few times $\sigma_{\rm c} m_e c^2$, as the cooling is inefficient in current sheets. However, once particles are captured by the plasmoids, where the magnetic field is high, and the cooling time is short, they rapidly lose their energies. Because of that we do not expect to observe any growth in particle cutoff energy with time for $\gamma_{\rm rad}\lesssim\sigma_{\rm c}$, as was predicted by \cite{2018arXiv180800966P} in the uncooled case. 
\end{itemize}

\begin{itemize}[leftmargin=*,wide = 0pt]
    \item For weak cooling, the peak in photon spectrum is set by the cutoff in particle spectrum (few times $\sigma_{\rm c}$). For strong cooling, the peak in photons corresponds to $\gamma_{\rm rad}$ in particle energy. Photons form $\nu F_{\nu}\propto \nu$ spectrum at low energies with a wide power-law tail at higher energies, close to $\nu F_{\nu}\propto \nu^{-1/2}$, up to a cutoff. 
    
    Photon spectrum beyond the peak is nonstationary with strong time variability, especially for stronger cooling regimes. Over time, these fluctuations add up to form a power-law tail. The time variability is primarily caused by plasmoid mergers, where particles are violently accelerated up to a few times $\sigma_{\rm c}$ in secondary current layers and cool down by radiating high energy photons when captured in merging plasmoids. 

\end{itemize}



\subsection{Two-photon pair production} 

Synchrotron photons, tracked in our simulations as regular massless and chargeless particles, can pair produce in the upstream and feed the current layer with secondary plasma. This process decreases the effective magnetization, suppressing the acceleration, and thus, the radiation and further pair production. 

\begin{itemize}[leftmargin=*,wide = 0pt]
    \item Pair production drives the system to a self-regulated steady state, where the initial upstream magnetization, $\sigma_{\rm c}$ is reduced by a factor of the resulting secondary plasma multiplicity.
    
    Simple analytical model for this steady state predicts the following relation for the multiplicity of secondary plasma near the light cylinder in $\gamma$-ray pulsars (see \S\ref{sec:theory}):
    \begin{equation}
        \eta_{\rm LC} \approx 180~\left(\frac{B_{\rm LC}}{10^5~\text{G}}\right)^{5/2}\left(\frac{P}{100~\text{ms}}\right)^{3/2}\frac{(s_1^2 s_2)^{1/3}}{0.1~R_{\rm LC}},
    \end{equation}
    where $B_{\rm LC}$ and $P$ are the magnetic field at the light cylinder and the period of the pulsar, and $s_1$ and $s_2$ are the sizes of the regions where most of the radiation and pair production take place. For the Crab pulsar this formula predicts $\eta_{\rm LC} \sim 10^4$.
\end{itemize}

\begin{itemize}[leftmargin=*,wide = 0pt]
    \item Pairs produced in the upstream form an extended power-law slope, which they inherit from their parent photons. This makes the inflowing secondary plasma hot. 
    
    This effect causes the particle energy distribution to depart from the standard $f(\gamma)\propto \gamma^{-1}$ spectrum to form a wide power law tail with indexes changing from $\gamma^{-2}$ at low energies to $\gamma^{-4}$ at higher energies. 
\end{itemize}

\begin{itemize}[leftmargin=*,wide = 0pt]
    \item Particles in the plasmoids radiate high energy photons that can pair produce in the peripheries of these plasmoids. This process feeds plasmoids with newly born secondary plasma and causes the plasmoids to rapidly inflate. These ``monster" plasmoids are typically larger and move slower than in the case of no pair production. They can capture a significant fraction of the simulation box, temporarily decreasing the reconnection rate. 
\end{itemize}



\begin{itemize}[leftmargin=*,wide = 0pt]
    \item In most of the runs with the radiation and pair production $\sim30-40\%$ of the total magnetic field energy in the box is deposited into particles and radiation equally (see details in the appendix~\ref{sec:appendixB}). Resistive MHD \citep{2012ApJ...749....2K} and PIC \citep{2014ApJ...785L..33P} simulations of the global magnetosphere predict that around $10\%$ of the total Poynting flux in pulsars is dissipated near the light cylinder. Combined with our finding of the efficiency of relativistic reconnection for generating radiation, this means that a few $\%$ of the spin-down energy is radiated as synchrotron radiation from the outer magnetosphere.
\end{itemize}

\begin{itemize}[leftmargin=*,wide = 0pt]
    \item By running simulations of reconnection with different magnetizations we find that $\gamma_{\rm cutoff}\propto B_0^{0.2\text{-}0.5}$, and the corresponding photon cutoff scales as $E_{\rm cutoff}\propto B_0^{1.2\text{-}1.8}$ with the background magnetic field, significantly weaker than without pair formation. 
    
    We did not vary the density of the inflowing primary plasma, which in the case of pulsars should scale with the corresponding Goldreich-Julian density near the light cylinder (thus with the background magnetic field near the light cylinder). This scaling would make the expected dependence on the magnetic field even weaker. 
\end{itemize}

\subsection{Observational implications}

In our simulations we demonstrated that the effects of two-photon pair production are crucial to consider even in the optically thin regime, when most of the high-energy photons leave unaffected by pair production. We have shown that this effect leads to the weak dependence of synchrotron spectrum cutoff in pulsars on the magnetic field strength near the light cylinder. In particular, observations with {\it Fermi Observatory} find $E_{\rm cutoff}\propto B_{\rm LC}^{0.1}\text{-}B_{\rm LC}^{0.2}$ (see Fig.~\ref{fig:fermi_data}). This cutoff energy is set by the maximum energy of accelerated particles, $\gamma_{\rm cutoff}$, and the background magnetic field near the light cylinder, i.e., $E_{\rm cutoff}\propto \gamma_{\rm cutoff}^2 B_{\rm LC}$. The particle energy cutoff is determined by the effective magnetization, $\gamma_{\rm cutoff}\sim \sigma_{\rm LC}\propto B_{\rm LC}^2 / \eta n_{\rm GJ}$, which is smaller by the factor of multiplicity of the secondary plasma produced near the light cylinder. \par

From our simulations with the constant inflowing plasma density (i.e., $n_{\rm GJ}$ does not vary with the magnetic field) we find that $E_{\rm cutoff}\propto B_{\rm LC}^{1.2}\text{-}B_{\rm LC}^{1.8}$. If we additionally assume that $n_{\rm GJ}\propto B_{\rm LC}$, we can infer that $E_{\rm cutoff}\propto B_{\rm LC}^{-0.8}\text{-}B_{\rm LC}^{-0.2}$. Our simulations, however, were only considering a two-dimensional isolated current sheet. For a more detailed application to $\gamma$-ray pulsars this effect needs to be studied more closely using the global magnetospheric simulations \citep{PSAS18}.

Our model suggests that pair production in the outer magnetosphere may efficiently increase the density of the plasma escaping to the pulsar wind nebula. Observed injection rate of $X$- and $\gamma$-ray emitting particles to the PWN suggests multiplicities $\sim 10^4$ with respect to the average Goldreich-Julian density. This number is consistent with the upper limit provided by pair creation models from primary cascade near the polar cap \citep{2018arXiv180308924T}. However, observations of radio-emission from low energy plasma population implies a much higher limit on the ultimate multiplicity at the level of $10^5\text{-}10^6$ for  several different nebulae \citep{2010ApJ...720..266S, 2011MNRAS.410..381B, 2012SSRv..173..341A}. In this paper we argue that on top of the primary cascade near the polar cap, two-photon pair production in the outer magnetosphere can further enhance the density of the outflowing plasma. This can account to the anomalously rich low-energy plasma population in the PWN observed in radio.


It is important to emphasize that in the present paper we studied the optically thin regime to both Thomson scattering and pair production, with only synchrotron radiation enabled, and with pair production driven by the interactions of these synchrotron photons. While this regime is applicable to pulsars, the process of two-photon pair production in reconnecting regions is ubiquitous and might be important in a wide range of astrophysical environments, in regimes where other QED mechanisms may also be important. We list some examples below.\par

\begin{itemize}[leftmargin=*,wide = 0pt]
    \item {\it Black hole accretion disk coronae} are thought to sustain reconnecting current layers with enough $\sim$MeV photons to trigger the production of secondary pairs. This effectively increases the Thomson optical depth of the layer to unity \citep{1983MNRAS.205..593G, 1987MNRAS.227..403S, 1993ApJ...413..507H}. In this optically thick environment both the synchrotron and inverse-Compton (IC) cooling as well as the $e^-e^+$ annihilation are important to consider \citep{2017ApJ...850..141B}. 
    \item {\it Blazar jet flares} with durations of several hours to days have been interpreted as powered by large and slow plasmoids in the reconnection layer, where two-photon pair production of IC photons may play a crucial role \citep{2012MNRAS.424L..26G, 2016MNRAS.462.3325P}. 
    \item {\it Gamma-ray flares in SGRs} are also thought to be powered by relativistic reconnection with a near-critical magnetic field and with ongoing pair production driven primarily by $\gamma+B$ process (see, e.g., \citealt{2001ApJ...561..980T, 2006MNRAS.367.1594L}, studied in particle-in-cell by \citealt{2018arXiv180709750S}) 
\end{itemize}

The approach introduced in this paper suggests a novel framework that could help to incorporate additional pair production effects in particle-in-cell simulations in a self-consistent way. We plan to continue improving our algorithms and investigating these unexplored regimes in future work.


We thank Dmitry Uzdensky for initial discussions on implementation of pair production in PIC codes, and Lorenzo Sironi for fruitful conversations. This work was supported by NASA grant NNX15AM30G, NASA through Einstein Postdoctoral Fellowship grant PF7-180165 awarded to AP by the Chandra X-ray Center, operated by the Smithsonian Astrophysical Observatory for NASA under contract NAS803060, Simons Foundation (grant 267233 to AS), and was facilitated by Max Planck/Princeton Center for Plasma Physics. The Flatiron Institute is supported by the Simons Foundation. The simulations presented in this article used computational resources supported by the PICSciE-OIT High Performance Computing Center and Visualization Laboratory and by NASA/Ames HEC Program (SMD-16-7816, SMD-17-455).

\appendix

\section{Details of the algorithm}

\label{sec:appendixA}

Since this is the first implementation of the self-consistent pair production in a particle-in-cell code, in this Appendix we present the details about the algorithm we used in our simulations. 

\begin{figure}[htb]
    \centering
    \includegraphics[width=.75\columnwidth]{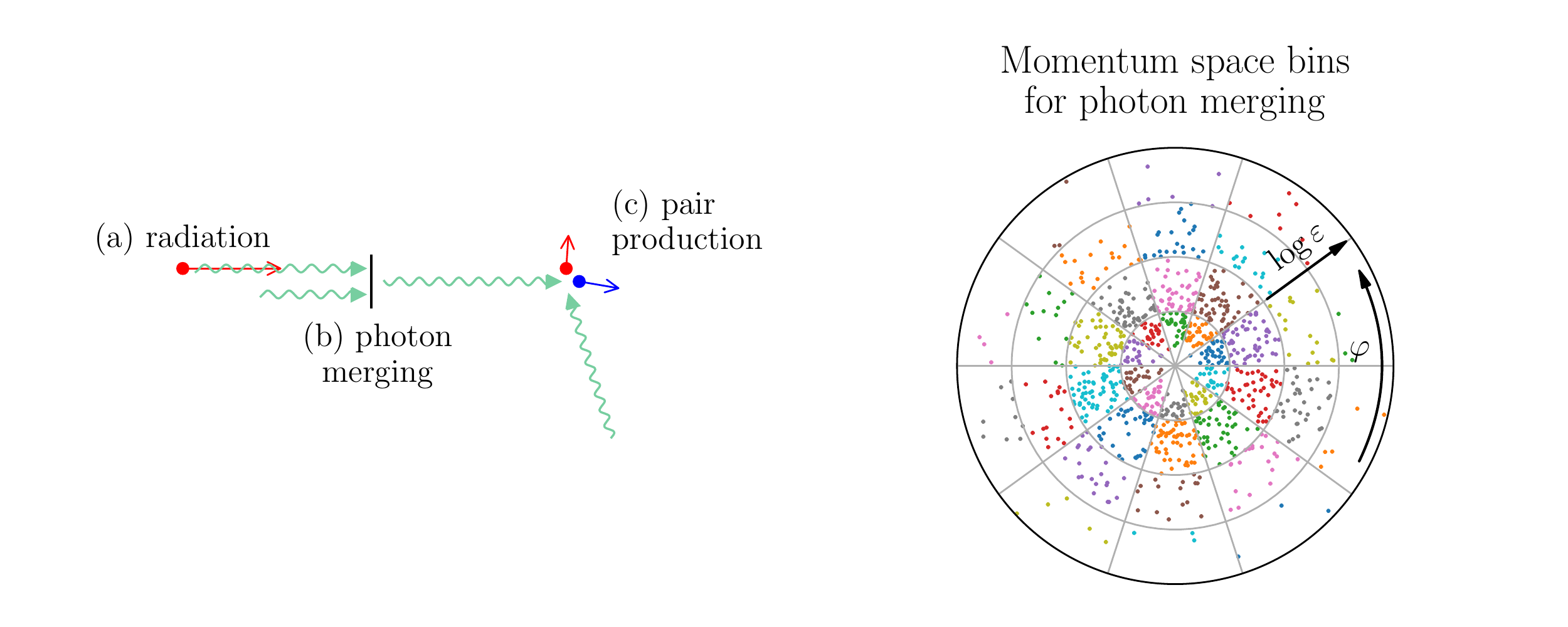}
    \caption{{\it Left}: three main steps in our algorithm -- plasma particle radiation a photon, photons merging (downsampling) in a given cell, photons interacting and producing secondary pairs. {\it Right}: two-dimensional version of the momentum binning we use for the photon downsampling. Binning is logarithmic in energy (momentum magnitude) and uniform in direction. Photons are colored according to their location in those bins, all the photons of the same color will be merged into two photons with higher weights conserving energy and momenta.}
    \label{fig:alg_scheme}
\end{figure}

We track two particle species: charged and massive plasma particles and massless photons. At each timestep a plasma particle can radiate a photon (step (a) in Figure~\ref{fig:alg_scheme}, {\it left}) with a corresponding synchrotron energy given by formula (\ref{eq:synch_form}), and the overall cooling rate is set by relation (\ref{eq:synch_rate}). The photons are then resorted in memory according to their spatial location.

Since we intend to study the optically thin regime to pair production, $\tau_{\gamma\gamma}\ll1$, and at the same time we have a sufficiently high multiplicity of secondary pairs, in our simulations the typical number of photons greatly exceeds the number of particles. This can very quickly exhaust the memory capabilities. To avoid this we use a downsampling (merging) algorithm for the photons similar to one described by \cite{2015CoPhC.191...65V}.

In each simulation cell we define three-dimensional photon momentum bins and sort photons according to their momenta as seen in Figure~\ref{fig:alg_scheme}, {\it right}. The bins are logarithmic in photon energies and uniform in 3D directions. We also randomly ``rotate'' the bins to minimize any downsampling artifacts. All the photons in the same momentum bin are then merged into two photons with higher effective weights (step (b) in Figure~\ref{fig:alg_scheme}, {\it left}), conserving total energy and momentum. Note also, that since the downsampling is done for the lowest energy photons (which are the majority) and for those who have small relative momenta angles, downsampling does not strongly affect the pair production efficiency, since those photons have a negligible probability to pair produce.

Two-photon pair production (step (c) in Figure~\ref{fig:alg_scheme}, {\it left}) is another expensive step that we implement in our simulation. At each cell we loop through all the non-repetitive pairs of photons and compute their binary probabilities $p_{ij}$ to pair produce given their energies, momenta and the cross section formula. 

Since the weights of those photons can be greater than one, these probabilities as well can exceed unity, i.e., if $p_{ij}=4.2$ on average from these photons $i$ and $j$ we will create $4.2$ $e^-e^+$ pairs: $4$ pairs with a probability $1$ and one more pair with a probability $0.2$ (reducing the photon weights each time). This approach is designed to mimic as if these interactions were between independent photons not merged into a two ``heavy'' ones. 

The probability magnitudes are normalized to a fiducial parameter, $p_0$, which is chosen to ensure the low optical depth to pair production. Overall the optical depth for a photon can be written as
\begin{equation}
    \tau_{\gamma\gamma} = L\langle n_{\rm ph}\rangle f_0 p_0,
\end{equation}
where $L$ is the effective size of the system, $\langle n_{\rm ph}\rangle$ is the average number of (potentially pair producing) photons per cell along the path, $p_0$ is our fiducial parameter, and the prefactor $f_0$ accounts for the cross section for different energies and momenta orientation (see Figure~\ref{fig:sigma_pp}) and is typically $0.1\text{-}0.01$. In our simulations the size of the system is typically a few times $10^3$ cells, and the effective number of pair producing photons along the path can vary $10^2\text{-}10^3$ (less than the total number of photons per cell). This gives us a rough estimate that
\begin{equation}
    \tau_{\gamma\gamma} \sim \frac{p_0}{10^{-3}}.
\end{equation}

\begin{figure}[tb]
    \centering
    \includegraphics[width=.9\columnwidth]{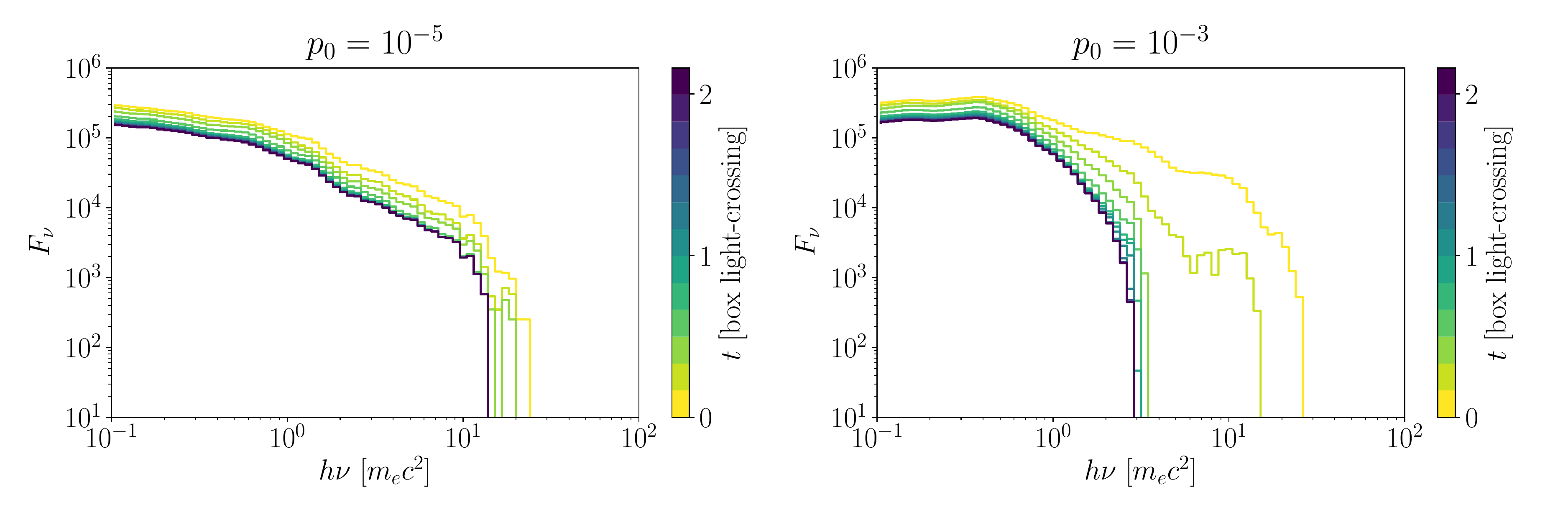}
    \caption{Time evolution of the spectrum of photons born at the same timestep. The time is measured in box light-crossing times, {\it left} plot corresponds to an optically thin regime with $p_0=10^{-5}$, and {\it right} -- optically thick with $p_0=10^{-3}$. Spectra are corrected to photons leaving the box, i.e., the reduction is only due to pair production. In optically thick regime most of the high energy photons interact in less than a light-crossing time.}
    \label{fig:opt_thick}
\end{figure}

The difference between optically thick and thin regimes is demonstrated in Figure~\ref{fig:opt_thick}. The evolution of a single photon generation spectra are different in these two cases ($p_0=10^{-3}$ and $p_0=10^{-5}$). In optically thick regime (Figure~\ref{fig:opt_thick}, {\it right}) most of the high energy photons interact with lower energy ones and pair produce in less than a single light-crossing time, resulting in a lower cutoff energy, whereas in optically thin regime (Figure~\ref{fig:opt_thick}, {\it left}) the spectrum nearly uniformly drops down over all energies due to pair production in a few light-crossing times.

\begin{table}
\begin{centering}
 \begin{tabular}{ l || c | c } 
 \hline
   & Computational cost & Memory usage \\ [0.5ex] 
 \hline\hline
 photon sorting & $\mathcal{O}(N_{\rm ph})$ & $\mathcal{O}(N_{\rm ph})$ \\
 photon merging & $\mathcal{O}(N_{\rm ph})$ & $\mathcal{O}(n_{\rm ph})$ \\
 pair production & $\mathcal{O}(n_{\rm ph} N_{\rm ph})$ & $\mathcal{O}(N_{\rm ph})$ \\ [1ex] 
 \hline
\end{tabular}
\caption{Time and memory costs for the most expensive parts of our algorithm as a function of the total number of photson in the box, $N_{\rm ph}$, and the average number of photons in each cell, $n_{\rm ph}$. }
\label{table:comp}
\end{centering}
\end{table}

Finally, in Tab.~\ref{table:comp} we present the time and memory consumption of our algorithm as a function of the total number of photons, $N_{\rm ph}$, and the average number of photons per simulation cell, $n_{\rm ph}$. Pair production is the most expensive procedure, since it is $\sim\mathcal{O}(n_{\rm ph}^2 N_{\rm cells}) \sim \mathcal{O}(n_{\rm ph} N_{\rm ph})$. 

Merging is efficient as far as the average number of photons per cell is $n_{\rm ph}\gg N_{\rm bins}$, where the number of momentum bins we typically use is $N_{\rm bins}=8^3=512$. In our typical run we have $10^4\text{-}10^5$ photons per cell, and, thus, this downsampling significantly decreases the cost by reducing the number of tracked photons typically by a factor of $10\text{-}100$.

In most of our runs this is still expensive, and we do this procedure once every several steps, instead of doing it every step. One, however, should keep in mind, that this interval cannot be longer than the typical mean free path of the photons to pair production (which in our case is a fraction of the box size), and also the interval should be short enough for the merging to prevent the memory exhaustion.

\section{Radiation and pair production statistics}
\label{sec:appendixB}

\begin{figure}[tb]
    \centering
    \includegraphics[width=0.5\columnwidth]{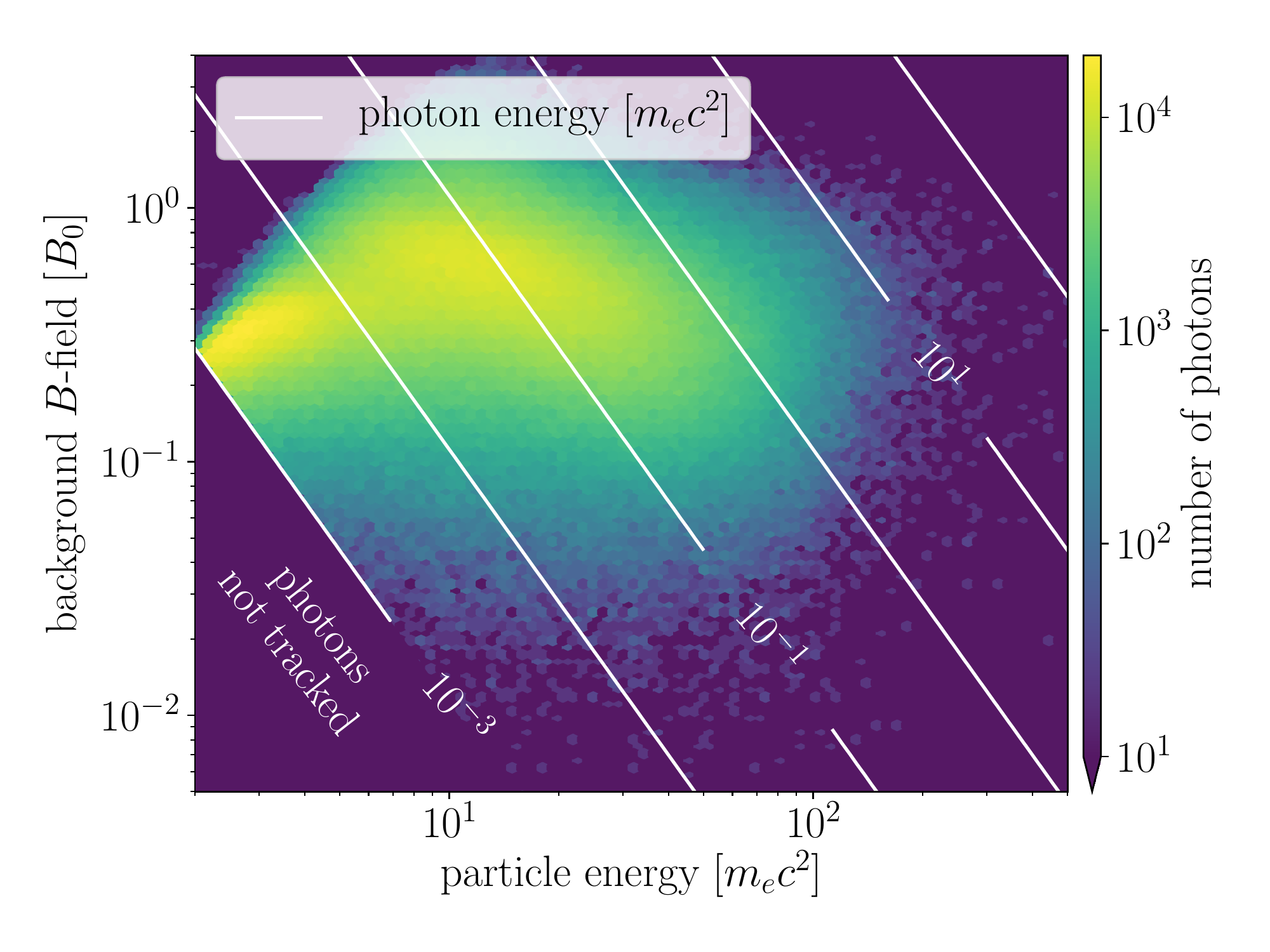}
    \includegraphics[width=0.4\columnwidth]{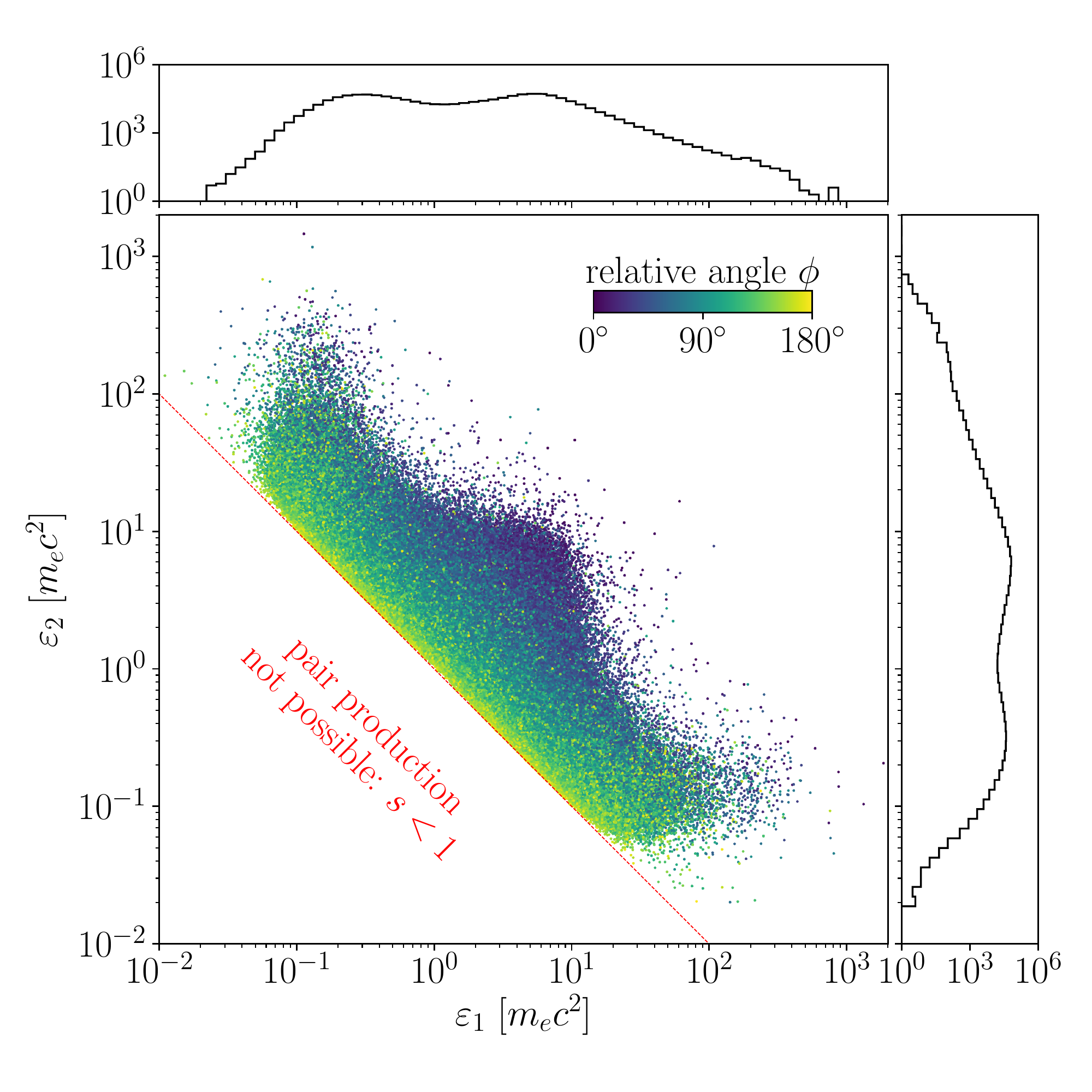}
    \caption{{\it Left:} 2D histogram of the radiated synchrotron photons vs the particle energy and the background magnetic field from one of our simulations ($\sigma_{\rm c}=5000$, $\gamma_c=50$, $\gamma_{\rm rad}=200$, $p_0=10^{-5}$). Each particle radiates at a corresponding synchrotron frequency (white contours), set by an effective background magnetic field (y-axis) and the Lorentz-factor of the particle (x-axis). Photons below $10^{-3}m_e c^2$ are not tracked, since they do not participate in pair production. {\it Right:} pair production statistics from the same simulation as on the left. Each point is a pair production event with $x$ and $y$ axes corresponding to the interacted photon energies and colors corresponding to their relative angles. One dimensional histograms are shown above and to the right, demonstrating that a wide range of photon energies are roughly equally important in terms of pair production.}
    \label{fig:phot_born}
\end{figure}

We also present several diagnostic plots to justify our assumptions made earlier. Figure~\ref{fig:phot_born} ({\it left}) shows the two-dimensional histogram of the number of produced synchrotron photons plotted against the plasma particle energy and effective magnetic field that a particle experienced when radiating. Contour lines show the corresponding synchrotron energies. 

One can see that most of the photons are produced in a narrow range of magnetic field values from $0.1B_0$ to $B_0$, and the range gets even smaller for the higher energy particles, which are interesting in terms of pair production. Also it is clear that most of the photons are very low energy, which, however, do not strongly contribute to pair production. Thus, it is important to correctly set the minimum tracking energy to make sure to capture enough pair production, but at the same time not to overwhelm the memory.

Figure~\ref{fig:phot_born} ({\it right}) shows the statistics of pair production from the same run, scatter plotted against the energies of two photons that produced the pairs. Each point is the pair production event, the color of each point represents the relative angle, $\phi$, between photon momenta. As one could have anticipated, the closer the energy product $\varepsilon_1\varepsilon_2$ to $2m_e c^2$, the closer the relative angle to $180^{\circ}$, and vice versa: two high energy photons can interact if their relative angle is small. 

As one can also see from the one-dimensional histograms (Figure~\ref{fig:phot_born}, {\it right}), the majority of pair production is for the photon energies $\varepsilon>10^{-2}m_e c^2$, and, thus, the photon tracking energy limit (which in this case is $10^{-3} m_e c^2$) is justified. Two extended scatter ``wings'' to the right and up are due to the fact that some very high energy photons do not have a low energy partner to interact (not tracked), and are left to interact with the higher energy ones. These tails, while being a numerical artifact, however, do not contribute much to pair production. We have carried convergence tests with lowering the energy limit with similar results: very low energy photons do not have a a significant contribution to pair production.

\begin{figure}[tb]
    \centering
    \includegraphics[width=0.9\columnwidth]{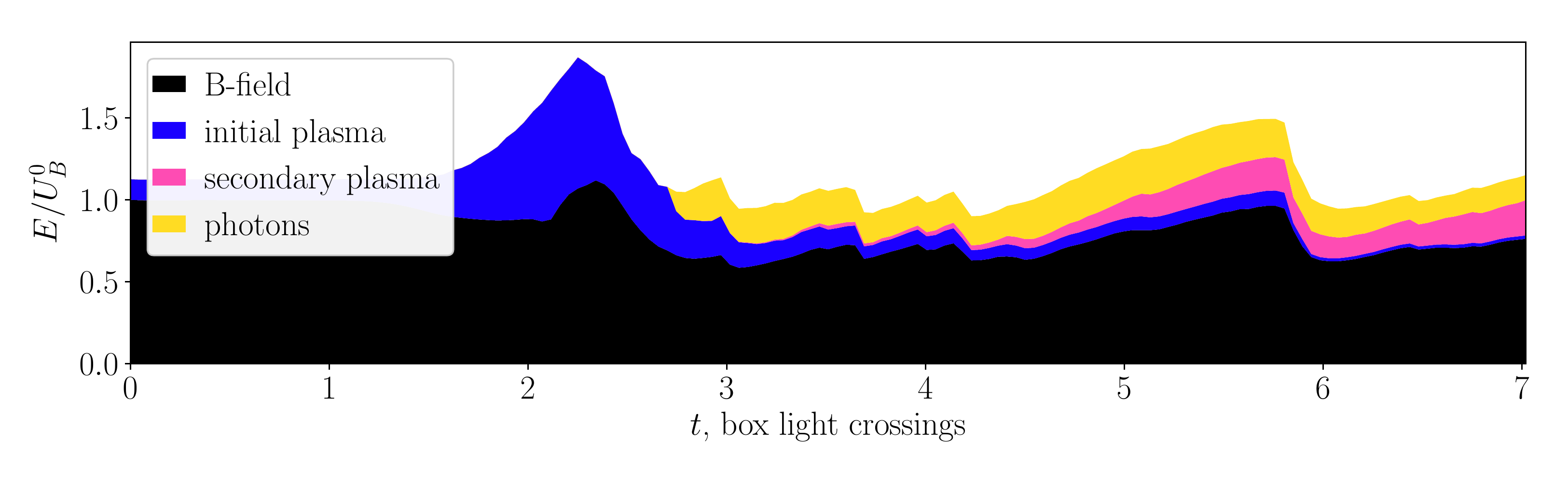}
    \caption{Energy partition in the simulation with $\sigma_{\rm c}=5000$, $\gamma_c=50$, $\gamma_{\rm rad}=1000$, $p_0=10^{-5}$. The energy is normalized by the initial magnetic field energy. Cooling and pair production is turned on at around three light-crossing times of the box. Large ``waves" are due to plasmoids forming, evolving and leaving the box in a few light-crossing times. One should keep in mind that the total energy of the magnetic field depends on the size of the region around the current sheet, and in this context it is rather artificial.}
    \label{fig:energy_vs_time}
\end{figure}

In Figure~\ref{fig:energy_vs_time} the distribution of total energy in the box between different components is shown (normalized by the initial magnetic field energy, $U_B$). Most of the energy is carried by the magnetic field which in the process of regular reconnection is being transferred to primary generation of particles up to 2-3 box light-crossing times. At that point the synchrotron cooling and pair production are turned on and the reconnection relaxes to a new steady state. 

At late times the energy is mostly contained in photons and secondary particles created in pair production events. The large ``waves" lasting a few box light-crossing times at late times are due to the large plasmoid, that constantly form, accumulate secondary plasma from the environment, and are then advected out from the box. 

\bibliography{references}

\end{document}